\documentclass[pdflatex,sn-apa]{sn-jnl}

\usepackage{graphicx,calc}%
\usepackage{graphicx}%
\usepackage{array}%
\usepackage{multirow}%
\usepackage{amsmath,amssymb,amsfonts}%
\usepackage{amsthm}%
\usepackage{mathrsfs}%
\usepackage[title]{appendix}%
\usepackage{xcolor}%
\usepackage{textcomp}%
\usepackage{manyfoot}%
\usepackage{booktabs}%
\usepackage{algorithm}%
\usepackage{algorithmicx}%
\usepackage{algpseudocode}%
\usepackage{listings}%
\usepackage{wrapfig}

\usepackage[utf8]{inputenc}%
\usepackage[T2A]{fontenc}%
\usepackage{caption} 
\usepackage[misc,geometry]{ifsym}
\usepackage{subcaption}
\usepackage[font=small]{caption, subcaption}
\usepackage[section]{placeins}
\usepackage{flushend}
\usepackage{pdflscape}
\usepackage{adjustbox}
\usepackage[english]{babel}

\theoremstyle{thmstyleone}%
\theoremstyle{thmstyletwo}%

\theoremstyle{thmstylethree}%

\begin{document}

\title[Article Title]{The Adaptive Strategies of Anti-Kremlin Digital Dissent in Telegram during the Russian Invasion of Ukraine}


\author[1]{\fnm{Apaar} \sur{Bawa}}\email{bawa.apaar@gmail.com}

\author[1]{\fnm{Ugur} \sur{Kursuncu}\textsuperscript{\Letter}}\email{ugur@gsu.edu}

\author[2]{\fnm{Dilshod} \sur{Achilov}}\email{dachilov@umassd.edu}

\author[3]{\fnm{Valerie} \sur{L. Shalin}}\email{valerie.shalin@wright.edu}

\affil[1]{\orgdiv{Georgia State University}, \state{Atlanta, GA}, \country{USA}}

\affil[2]{\orgdiv{University of Massachusetts}, \state{Dartmouth, MA}, \country{USA}}

\affil[3]{\orgdiv{Wright State University}, \state{Dayton, OH}, \country{USA}}


\abstract{During Russia's invasion of Ukraine in February 2022, Telegram became an essential social media platform for Kremlin-sponsored propaganda dissemination. Over time, Anti-Kremlin Russian opposition channels have also emerged as a prominent voice of dissent against the state-sponsored propaganda. This study examines the dynamics of Anti-Kremlin content on Telegram over seven phases of the invasion, inspired by the concept of breach in narrative theory. A data-driven, computational analysis of emerging topics revealed the Russian economy, combat updates, international politics, and Russian domestic affairs, among others. Using a common set of statistical contrasts by phases of the invasion, a longitudinal analysis of topic prevalence allowed us to examine associations with documented offline events and viewer reactions, suggesting an adaptive breach-oriented communications strategy that maintained viewer interest. Viewer approval of those events that threaten Kremlin control suggests that Telegram levels the online playing field for the opposition, surprising given the Kremlin’s suppression of free speech offline.}


\maketitle

\section{Introduction}
\label{sec1}

“Should we just turn the world to dust?” asked Vladimir Solovyov during his live show on April 29, 2022, threatening the international community with the potential use of nuclear weapons \citep{Gessen2022}. This was not the first or last of such escalatory rhetoric from  Solovyov. In fact, with his show airing six times a week in Russia, Mr. Solovyov is probably the most prominent and prolific pro-Kremlin propagandist of all time \citep{USDOS2022}. In addition to his live TV and radio shows, Solovyov’s Telegram social media channel, which has over 1.4 million subscribers, has long helped to shape public discourse in the Russian political landscape. Since 2005, the Russian government, led by President Vladimir Putin, has utilized different online platforms to disseminate propaganda and disinformation, generate support for his domestic and foreign policies, stifle pro-democracy voices through systematic restrictions, and conduct mass surveillance of Russian citizens \citep{Sanovich2017}. No doubt, the Kremlin has made a systematic effort to consolidate media under state control. Major media outlets (e.g., newspapers, TV channels) in Russia, like Channel One, Rossiya 1, and NTV, which reach most of the population, are either state-owned or controlled by entities with close connections to the Kremlin \citep{Hutchings2009}. This enables the Kremlin to have significant influence in shaping the narratives presented to the public with favorable sentiments of the state. The Kremlin's influence on media extends beyond its borders. This includes discussion surrounding Russia's long-standing military support in the Syrian civil war since 2012 \citep{Strovsky2020} and the Russian military aggression against Ukraine in February 2022 \citep{Hanley2023}. Countering this influence is a daunting task. \bigskip

Nevertheless, alternative social media platforms have emerged as crucial spaces for independent and dissenting voices \citep{Enikolopov2020}. Anti-Kremlin Telegram channels challenge the state-sponsored official accounts of events, mobilizing public opinion against the Kremlin’s “special operations” narrative. Over time, these channels have evolved from mere news aggregators to influential voices of dissent, exposing government wrongdoings, promoting democratic values, and advocating for change \citep{Castells2015}. Here, we examine the Anti-Kremlin communication as strategic, not in controlling events for which they are patently unable, but rather, a media-empowered structuring and distribution of event representations that constitute an alternative longitudinal narrative. First, we document a reduction of content in response to an anticipated event that failed to materialize. Second, we demonstrate an increase in content tethered to real-world events. Third and most generally, we demonstrate that the Kremlin has lost control over opposition channel content as measured by public engagement. \bigskip

Many researchers claim that we organize experience in temporal sequence, which we call a story. Narrative is a particular accounting, often employing a familiar genre such as romance or tragedy in the selection of notable detail and agent roles and intent \citep{Propp1968}.  Narrative researchers typically emphasize structural properties, for example, the juxtaposition of stability, disruption and resolution \citep{Todorov1969}. Bruner’s \citep{Bruner1991} breach theory provides a content-oriented account of stability and disruption in terms of social norms embedded in routine, canonical structures. When one of these structures experiences a “breach,” such as encountering new information that contradicts our established beliefs or expectations, it has the potential to challenge and even transform our cognitive perceptions \citep{Bruner1990}. Thus, we suggest that Anti-Kremlin Telegram endeavors to create a “breach” in the existing beliefs or expectations of a public otherwise entrapped by Pro-Kremlin propaganda, thereby undermining “manufactured consent” \citep{Herman2021}. Such breach creation efforts elicit a range of responses, from disregard (especially given a well-documented my-side bias \citep{Stanovich2013} in the interpretation of incoherent information) to belief modification, influenced by the breach’s nature and the online community’s standards \citep{Garfinkel1967}. \bigskip

We situate breach exploitation in the emerging theory of strategic communication. 
\citet{Hallahan2007} define strategic communication “in its broadest sense” as “the purposeful use of communication by an organization to fulfill its mission” (p. 3). Anti-Kremlin Telegram has a clear self-stated mission. Hallahan et al. further argue that strategic communication involves deliberate communication practices on behalf of organizations, causes, and social movements (p. 4), in our case, Kremlin resistance. 
\citet{Holtzhausen2014} suggests that strategic communication is fundamental to the source’s existence. Moreover, according to Holtzhausen \& Zerfass, “\textit{communication is not strategic …. when it is about known operational and routine issues} with well-established tactics of intervention” (p. 11 italics added for emphasis). \bigskip

Clarity and accuracy are not necessarily a requirement for strategic communication \citep{Dulek2015}, but capturing viewer attention is \citep{Brady2020}. 
\citet{Jarzabkowski2007} argue for the importance of what the communications literature refers to as \textit{audience design} \citep{Bell1984}, targeting specific recipients with context-sensitive relevant information \citep{Sperber1986}. However, relevance in a dynamic environment and repetition would merely create habituation \citep{Jankowski2021}, no matter how accurate. The need to garner and retain recipient attention in a dynamic environment mostly outside the organization’s control implies an adaptive requirement, sensitive to what makes something worth stating and noting. Breach adds to the strategic communication heuristics that focal actors can employ with social media in a dynamic environment. In social media, viewer reactions measure their attention, providing feedback to the source necessary for this adaptive capability \citep{kramer2021feel,kursuncu2019predictive}. We should see, therefore, a distinctive “strategic learning curve” whereby Anti-Kremlin channels recalibrate their messaging on Telegram dynamically, effectively countering the framing effects of the Kremlin’s communications. \bigskip

Our study examines over a million posts from 114 influential Anti-Kremlin communication channels on Telegram, from before the invasion of Ukraine to the protracted conflict that followed. The extended duration of the same overarching event with the same participants allows us to establish adaptive communication processes. Through an in-depth analysis of these Telegram posts in the timeframe from January 1, 2022, to March 31, 2023, we investigate the interplay of content production, user engagement, and key offline events. \bigskip

Our argument employs a quantitative analysis of Anti-Kremlin content with respect to the temporal phases of the Ukraine conflict, delineating one pre- and six post-invasion periods,  consistent with 
\citet{Murauskaite2023}. We collected posts from the 614 most popular Russian Anti-Kremlin political channels, manually labeled based on their content's alignment with the overall Kremlin rhetoric. We performed topic modeling for each temporal phase, utilizing the multilingual large foundation model MPNet \citep{Song2020} followed by BERTopic \citep{Grootendorst2022}. Then, we annotated the resulting topical clusters using the Gioia method \citep{Gioia2013}, identifying 356 low-level topics and 55 higher-level abstractions of these. To argue that topic trends are phase-sensitive, we use a common set of contrasts with post volumes with measures pertaining to different topics and engagement over the conflict phases. The different patterns of contrast significance by topic across phases support our claim that topics are adapting to context. Complementary event analysis provides insights into the evolving dynamics of topic change. We make three points regarding this adaptive capability: 

(i) An anticipated breach in economic security, initially emphasized by both the Anti-Kremlin channels and Western media, never materialized and is reflected in the initial elevation and subsequent reduction in Telegram posts and viewer reactions. 

(ii) A differing pattern of topic elevation and reduction demonstrates the general responsiveness of the Anti-Kremlin posts regarding international politics and combat updates, reinforced by viewer reaction, demonstrating an adaptive strategic capability.

(iii) The strategy effectively diminishes the Kremlin’s effort to control public opinion. 


\section{Results}
\label{sec2}

We first present initial results regarding the topic analysis. Then, we address the three main points regarding the economic argument, the event-specific responses, and the Kremlin’s loss of control over the narrative.

\subsection{Results from Topic Analysis}\label{subsec2}

\begin{table}[h!]
\centering
\begin{tabular}{p{2cm}p{2cm}p{7.8cm}}
\toprule[2pt]
\textbf{Categories} & \textbf{Subcategories} & \textbf{Keywords \footnotesize (English Translation)} \\ 
\midrule[1pt]
\multirow{3}{5em}{\parbox{1\linewidth}{\vspace{1.3cm} Russian Domestic Affairs}} 
    & Opposition Crackdown, Suppression of free speech & court, case, accusation, fake, kidnapping, activist, detain, arrest, torture, convict, hunger strike, antiwar, rally, foreign, agent \\ \cmidrule(lr){2-3} 
    & \vspace{0.15cm}Mobilization & containment, special, order, transfer, mobilization, partial, military, protest, detain, protest, rally, police, war, military registration and enlistment office, PMC, resignation, minister \\ \cmidrule(lr){2-3} 
    & Referendum & referendum, governor, elections, resignation, vote, annexation, municipal, party, september, territory, occupy \\ 
\midrule[1pt]
\multirow{3}{3em}{\parbox{1\linewidth}{\vspace{1.2cm} Economy}} 
    & \vspace{0.1cm}Currency & ruble, dollar, euro, fall, stock, bargaining, bank, default, moscow exchange, historical, swift, shutdown, inflation, GDP, a crisis, default, inflation \\ \cmidrule(lr){2-3} 
    & Gas/Oil & gas, supply, price, exceed, ceiling, pipeline, flow, stop, contract, payment, oil, import, embargo, export, leak \\ \cmidrule(lr){2-3}
    & Food export & grain, wheat, food, port, export, UN, export, global, ship, Turkey, Odessa, corridor, Black Sea, deal, extension \\ 
\midrule[1pt]
International Politics
    & - & Mentioned countries/topics: Kazakhstan, Georgia, Europe Visa, China, USA, Iran, Belarus, Turkey, Azerbaijan, Armenia, Israel, Moldova \\ 
\midrule[1pt]
\multirow{2}{5em}{\parbox{1\linewidth}{\vspace{0.3cm} Ukrainian Domestic Affairs}} 
    & Electricity shutdown & bridge, Crimean, included, shutdown, electricity, explosion, electricity, fire, light, Ukrenergo \\ \cmidrule(lr){2-3}
    & Refugee mobilization & refugee, exchange, Vershchuk, Ukrainian, captive, leave, million, citizenship, passport, refugee, visa, decree, UN \\ 
\midrule[1pt]
Combat and Frontline updates 
    & - & Russian, military, rocket, tank, occupants, Armed Forces of Ukraine, enemy, drone, attack, airplane, shoot, shelling, residential, war, hit, missile, perish, booth, himars, eagle, Bucha \\ \bottomrule[2pt]
\end{tabular}
\caption{\small Main results from the topic analysis, presenting the most prevalent five categories, their subcategories, and keywords. Subcategories are listed for Russian Domestic Affairs, Economy, and Ukrainian Domestic Affairs. International politics lacks subcategories because each country has unique relational keywords. Keywords were translated into English for better accessibility.}
\label{Table 1}
\end{table}

\begin{table}[h!]
\centering
\begin{tabular}{p{0.7cm}p{3.5cm}p{7.8cm}}
\toprule[2pt]
\textbf{Phase} & \textbf{\textit{Online Themes}} & \textbf{Key Offline Events} \\ 
\midrule[1pt]
1 & \textit{Economy} & Imposed sanctions and its anticipated impact on the Russian economy (\textit{Event A}) \\ 
\midrule[1pt]
2 & \textit{War and Combat updates} & Increase in intensity of war after Ukrainian counteroffensive with Western support (\textit{Event B}) \\ 
\midrule[1pt]
\multirow{2}{1em}{\parbox{1\linewidth}{\vspace{0.3cm}3}} & \vspace{0.1cm}\textit{Economy} & Russia defaulted on its external sovereign bonds for the first time in a century and billion dollars in aid to Ukraine (\textit{Event C}) \\ 
\midrule[1pt]
\multirow{2}{1em}{\parbox{1\linewidth}{\vspace{0.3cm}4}} & \vspace{0.1cm}\textit{Russian domestic affairs} & Declare partial mobilization in the Russian Federation/Russian annexation of occupied Ukrainian territory including Donbas (\textit{Event D}) \\ 
\midrule[1pt]
\multirow{2}{1em}{\parbox{1\linewidth}{\vspace{0.3cm}5}} & \vspace{0.1cm}\textit{Ukrainian domestic affairs} & Russian forces launched a massive wave of strikes against critical Ukrainian infrastructure in retaliation over the destruction of Kerch Strait bridge (\textit{Event E}) \\ 
\midrule[1pt]
\multirow{2}{1em}{\parbox{1\linewidth}{\vspace{0.8cm}6}} & \vspace{0.1cm}\textit{International politics} & Joe Biden visited Kyiv, Xi Jinping visited Russia, Iranian Foreign Minister met Russian Foreign Minister, and \$350 million of security assistance to Ukraine by the US (\textit{Event F}) \\ \cmidrule(lr){2-3}
 & \textit{War and Combat updates} & Again, increase in the intensity of war in relation to Event F (\textit{Event G}) \\ 
\bottomrule[2pt]
\end{tabular}
\caption{\small Qualitative mapping between prominent online themes and corresponding key offline events across conflict phases. The online discussions within Anti-Kremlin channels were responsive and dynamically changing, adapting to the ongoing offline, real-world events, with users contributing to public sentiment.}
\label{Table 2}
\end{table}

While our analysis identified 55 topical primary categories and 356 subcategories spanning the six post-invasion and pre-invasion, our main focus was on the five most prominent and persistent themes throughout these phases: Russian Domestic Affairs, Economy, International Politics, Ukrainian Domestic Affairs, and Combat and Frontline Updates. For these major themes, we identified second-order key subcategories. Russian Domestic Affairs included topics of Opposition Crackdown, Mobilization, and Referendum; the Economy contained Currency, Gas/Oil, and Food Export; International Politics covered Kazakhstan, Europe Visa, China, USA, Iran, and Belarus; Ukrainian Domestic Affairs included Electricity shutdown, and Refugee mobilization. The topical categories of International politics, and Combat and Frontline updates remained at a high level with coarse granular details; hence, without any subcategories. Table \ref{Table 1} presents these topical categories and subcategories with related keywords in a structured manner. These categories are formed based on TF-IDF keywords by context. Hence, the keywords appearing in multiple categories (e.g., military) have different semantics due to their varying contexts. An area chart of absolute post volumes and details on these themes with their original keywords in Russian are presented in Appendix Figure \ref{Appendix: Figure B1} and Appendix Table \ref{Appendix: Table A2}.

\begin{figure}[h!]
    \centering
    \includegraphics[width=\linewidth]{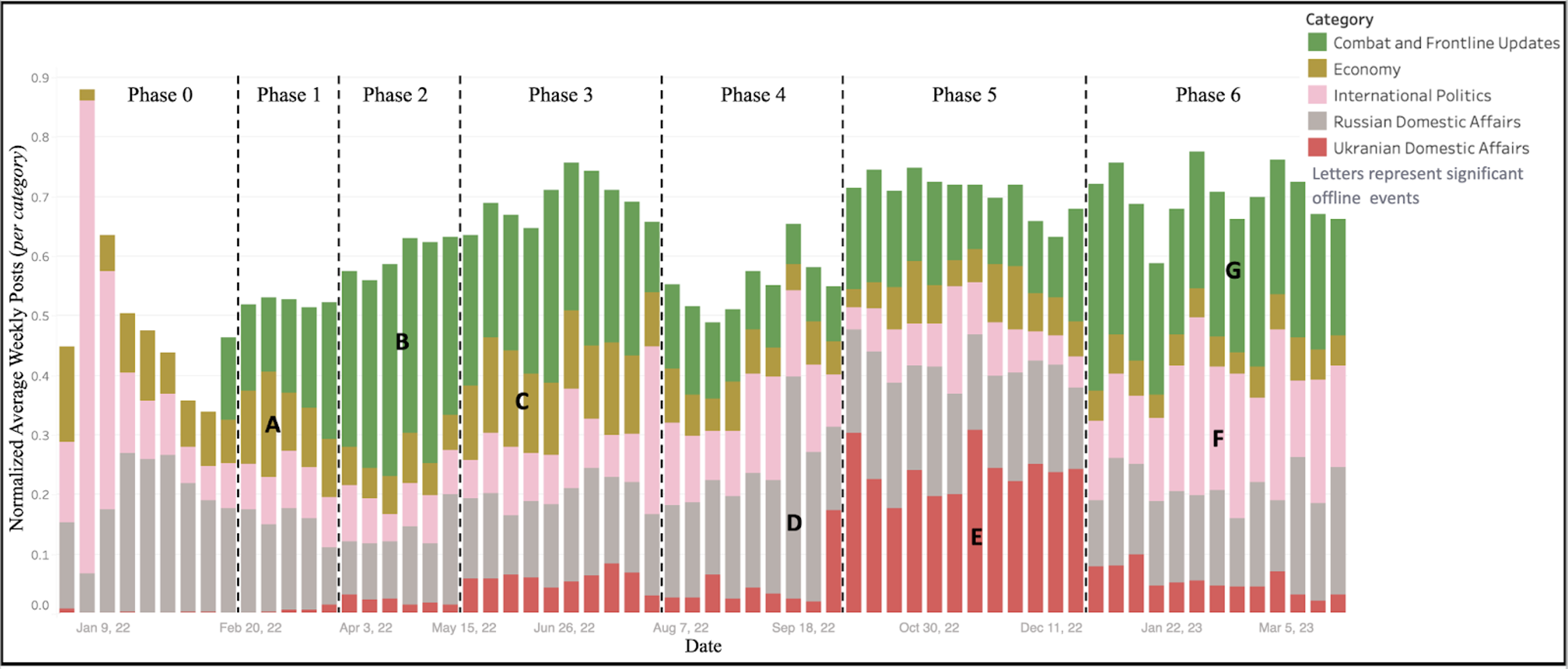}  
    \caption{Normalized \textit{weekly post volume (frequency)} \textit{Timeline (per each category)}. The bar chart represents the evolution of the top 5 categories over seven phases on a weekly basis. \textit{Categories are represented as Combat and frontline updates, Economy, International Politics, Russian domestic affairs, and Ukrainian domestic affairs. To compare the categories on the same scale within a week, we normalize them as a proportion between 0 and 1. Posts frequency Timeline, area chart represents the evolution of the top 5 categories over seven phases. Categories are represented as Combat and frontline updates, Economy, International Politics, Russian domestic affairs, and Ukrainian domestic affairs.} Major offline events are indicated by letters described in the text.}
    \label{fig:timeline}
\end{figure}

We mapped the dynamic relationship between these online discussions in Anti-Kremlin channels and corresponding offline events. Figure \ref{fig:timeline} illustrates the evolution and trends of key topical categories throughout the conflict's phases \citep{Murauskaite2023}, providing a weekly breakdown of post volumes per category, to highlight correspondence with key offline events. In this figure, min-max normalization was applied to the post volumes (absolute frequency counts) across the weeks throughout the phases to facilitate the visual identification of relative changes independent of fluctuations in the overall volume. \bigskip 

In Figure \ref{fig:timeline}, each online theme is distinguished by color, with specific co-occurring offline events (see Table \ref{Table 2}) marked from A to G. The economic sanctions (Event A) during Phase 1 occurred in response to the full Russian invasion (see Appendix Table \ref{Appendix: Table A3}). Phase 2, continuing into Phase 3, was marked with increased combat intensity (Event B), fueled by Western support for a Ukrainian counter-offensive. Event C highlighted Russia’s financial issues emerging upon Russia defaulting on its international bonds on June 27, 2022, attributed to the sanctions that isolated Russia from the global financial system, rendering its assets inaccessible, aiming to impact its economic stability \citep{Strohecker2022}. In Phase 4, the Kremlin declared partial mobilization and annexation of occupied Ukrainian territory (Event D).  Phase 5 concerns massive air strikes against Ukrainian infrastructure (Event E). Phase 6 coincided with the escalated Western military aid and intelligence support, boosting the Ukrainian resistance (Event F) and a subsequent Ukrainian counteroffensive, leading to the escalation in the intensity of war (Event G). 

\subsection{Results from Statistical Analysis}\label{subsec2}
\begin{figure}[h!]
    \centering
    \includegraphics[width=\linewidth]{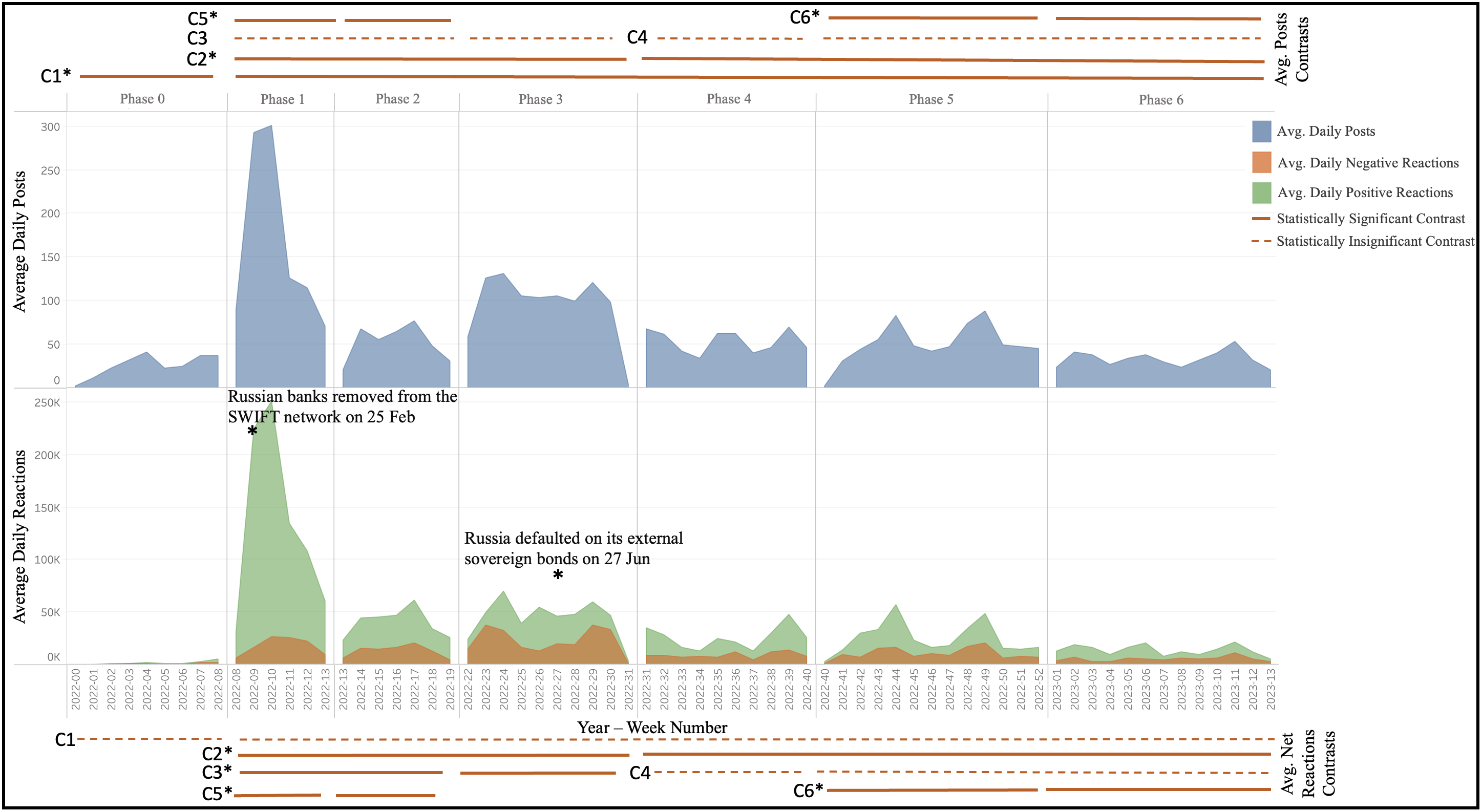}  
    \caption{Discussion of the Russian Economy across the phases. (i) in the top graph,  an analysis of changes in weekly average post volume (blue); (ii) in the bottom graph, the analysis of changes in weekly average positive (green) and negative (red) viewer reactions; (iii) top and bottom statistical significance (illustrated with solid lines) of post volumes and net viewer reactions, with contrasts across different phases (alpha <.05). We see a statistically significant surge in Anti-Kremlin channel posts related to economic sanctions (upper half)  and user engagement (lower half). An early bump, coincident with the removal of Russian banks from the SWIFT network, is apparent in contrasts 1, 2, and 5. Viewer reaction mirrors this bump in contrasts 2 and 5. Except for significant contrast 6, Anti-Kremlin posts have abandoned the Economic argument against the Kremlin.}
    \label{Fig:Economy}
\end{figure}

\begin{figure}[h!]
    \centering
    \includegraphics[width=\linewidth]{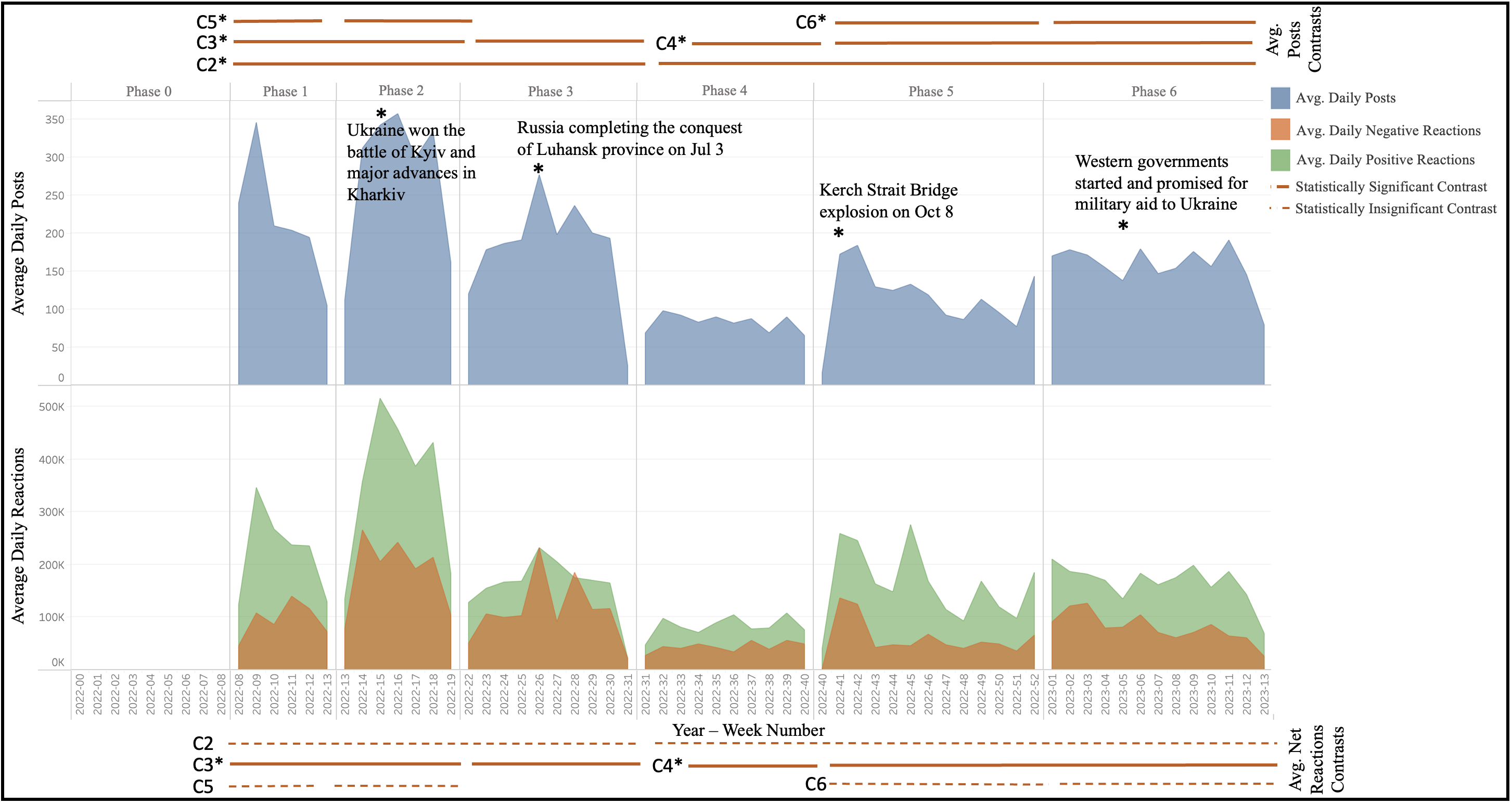}  
    \caption{Combat and Frontline Updates theme across the phases. We captured the (i) evolution of average weekly posts (blue area chart), (ii) evolution of positive (green) and negative (orange) user engagement, (iii) statistical significance of contrast weights for post counts and net reactions (illustrated with single solid lines/dotted lines). Phase 0 is not included as there is no data available; hence, Contrast 1 is omitted.}
    \label{Fig:Combat}
\end{figure}

To complement the qualitative impressions in Figure \ref{fig:timeline}, we employed the daily average number of posts and viewer reaction measures to an ANOVA framework with a common set of a-priori contrasts, treating phases as an-event based nominal variable. Our findings are encapsulated by the results of the three analyses, illustrated in Table \ref{Table 3}, Appendix Table \ref{Appendix: Table A1}, and Figures \ref{Fig:Economy}, \ref{Fig:Combat}, \ref{Fig:International_Politics}, and \ref{Fig:Russian_Dom_Aff}.

\newgeometry{margin=3cm,footskip=3em,}
\begin{landscape}
\small
\begin{table}[h!]
\centering
\begin{tabular}{p{1.2cm}p{1.25cm}p{0.9cm}p{0.9cm}p{0.9cm}p{0.9cm}p{0.9cm}p{0.9cm}p{0.9cm}p{0.9cm}p{0.9cm}p{0.9cm}p{0.9cm}p{0.9cm}p{0.9cm}p{0.9cm}p{0.9cm}p{0.9cm}}
\toprule[2pt]
\textbf{Contrasts} & \textbf{Factor Variable} & 
\multicolumn{4}{c}{\textbf{Economy}} & 
\multicolumn{4}{c}{\textbf{International Politics}} & 
\multicolumn{4}{c}{\textbf{Combat and Frontline Updates}} & 
\multicolumn{4}{c}{\textbf{Russian Domestic Affairs}} \\ 
    \cmidrule(r){3-18}
    & \tiny(avg daily per week) & \footnotesize\textbf{t-stat} & \textbf{p} & \textbf{dof} & \textbf{d} & \footnotesize\textbf{t-stat} & \textbf{p} & \textbf{dof} & \textbf{d} & \footnotesize\textbf{t-stat} & \textbf{p} & \textbf{dof} & \textbf{d} & \footnotesize\textbf{t-stat} & \textbf{p} & \textbf{dof} & \textbf{d} \\ 
\midrule[1pt]
\multirow{2}{5em}{C1: (P0 vs rest)} 
    & posts & -1.996 & 0.050* & 64 & -0.753 & 0.247 & 0.806 & 64 & 0.093 & - & - & - & - & -2.012 & 0.048* & 64 & -0.759 \\ \cmidrule(lr){3-18}
    & reactions & -1.936 & 0.057 & 64 & -0.730 & -8.745 & 3.76E-12* & 57.505 & -3.298 & - & - & - & - &  -7.549 & 3.45E-10* & 58.147 & -2.847 \\ 
\midrule[1pt]
\multirow{2}{5em}{C2: (P. 1-3 vs 4-6)} 
    & posts & 4.935 & 5.70E-05* & 22.668 & 1.336 & -0.448 & 0.656 & 56 & -0.121 & 8.002 & 1.03E-08* & 28.01 & 2.197 & 1.104 & 0.274 & 56 & 0.299 \\ \cmidrule(lr){3-18}
    & reactions & 3.568 & 0.002* & 21.636 & 0.966 & 2.521 & 0.015* & 56 & 0.682 & 2.014 & 0.054 & 27.959 & 0.553 & 1.3 & 0.199 & 56 & 0.352 \\ 
\midrule[1pt]
\multirow{2}{5em}{C3: (P. 1,2 vs 3)} 
    & posts & 0.502 & 0.621 & 20 & 0.218 & 1.516 & 0.145 & 20 & 0.657 & 3.16 & 0.005* & 19 & 1.394 & 1.486 & 0.153 & 20 & 0.644 \\ \cmidrule(lr){3-18}
    & reactions & 2.494 & 0.022* & 20 & 1.082 & 3.304 & 0.005* & 15.269 & 1.433 & 4.412 & 0.0003* & 19 & 1.946 & 4.234 & 0.0007* & 15.525 & 1.836 \\ 
\midrule[1pt]
\multirow{2}{5em}{C4: (P. 4 vs 5,6)} 
    & posts & 1.983 & 0.056 & 34 & 0.738 & 1.399 & 0.171 & 34 & 0.521 & -7.69 & 2.41E-08* & 27.694 & -2.862 & 1.748 & 0.089 & 34 & 0.651 \\ \cmidrule(lr){3-18}
    & reactions & 1.504 & 0.142 & 34 & 0.560 & 0.827 & 0.414 & 34 & 0.308 & -4.026 & 0.0003* & 34 & -1.498 & 5.369 & 5.69E-06* & 34 & 1.998 \\ 
\midrule[1pt]
\multirow{2}{5em}{C5: (P. 1 vs 2)} 
    & posts & 3.988 & 0.010* & 5.138 & 2.219 & 3.441 & 0.006* & 11 & 1.914 & -4.239 & 0.002* & 10 & -2.482 & 2.288 & 0.043* & 11 & 1.273 \\ \cmidrule(lr){3-18}
    & reactions & 4.186 & 0.002* & 11 & 2.329 & 6.525 & 4.28E-05* & 11 & 3.630 & -1.311 & 0.219 & 10 & -0.768 & 2.479 & 0.031* & 11 & 1.380 \\ 
\midrule[1pt]
\multirow{2}{5em}{C6: (P. 5 vs 6)} 
    & posts & 3.201 & 0.004* & 24 & 1.256 & -3.398 & 0.002* & 24 & -1.333 & -2.959 & 0.007* & 24 & -1.161 & 3.837 & 0.001* & 24 & 1.505 \\ \cmidrule(lr){3-18}
    & reactions & 2.534 & 0.023* & 15.202 & 0.994 & -2.557 & 0.017* & 24 & -1.003 & 1.705 & 0.101 & 24 & 0.669 & -1.537 & 0.137 & 24 & -0.603 \\ 
\bottomrule[2pt]
\end{tabular}
\caption{Statistical comparisons across different phases for various topics. Factor Variables, post volume and net reactions, are reported based on average daily values per week. Asterisks (*) indicate statistically significant differences (p < 0.05). P:Phase, C: Contrast, p: p value, dof: degrees of freedom, d: Cohen’s d.}
\label{Table 3}
\end{table}

\end{landscape}
\restoregeometry

\subsubsection{Finding 1: The Russian Economic Breach that Never Was}\label{subsubsec2}
Figure \ref{Fig:Economy} examines the Anti-Kremlin posts regarding the Russian economy across the timeline of the conflict. The widely anticipated collapse of the Russian economy due to Western sanctions \citep{Sonnenfeld2022}, did not dominate online communications beyond the early phases. This conclusion results from a significant (declining) C5 regarding post volumes. C2 and C6 are also significant, although these later phases of interest are muted relative to the early phases. An additional post-hoc Tukey’s test suggested that the differences between phases 3 and phase 4 were statistically significant (meandiff=-53.8; p-adj=0.0015; lower=-92.5326; upper=-15.0674). However, phases 4 and 5 were not statistically significant (meandiff=-2.8154; p-adj=1.0; lower=-38.2733; upper=32.6425), reinforcing the visible declining focus on the \textit{Russian Economy} as the conflict progressed. Audience reactions in the bottom graph generally support this pattern, regarding C2, C5, and C6, suggesting viewer approval regarding an Anti-Kremlin event. C2 documents the reduction of interest. These observations align with findings by Egorov et al. (2023), which suggest that the \textit{Russian Economy} demonstrated resilience against Western sanctions, including trade restrictions, asset freezes, and financial market exclusions, perhaps with later signs of weakness \citep{Kantchev2023}.

\subsubsection{Finding 2: Anti-Kremlin Communications are Adaptive to Emergent Breaches}\label{subsubsec2}

\begin{figure}[h!]
    \centering
    \includegraphics[width=\linewidth]{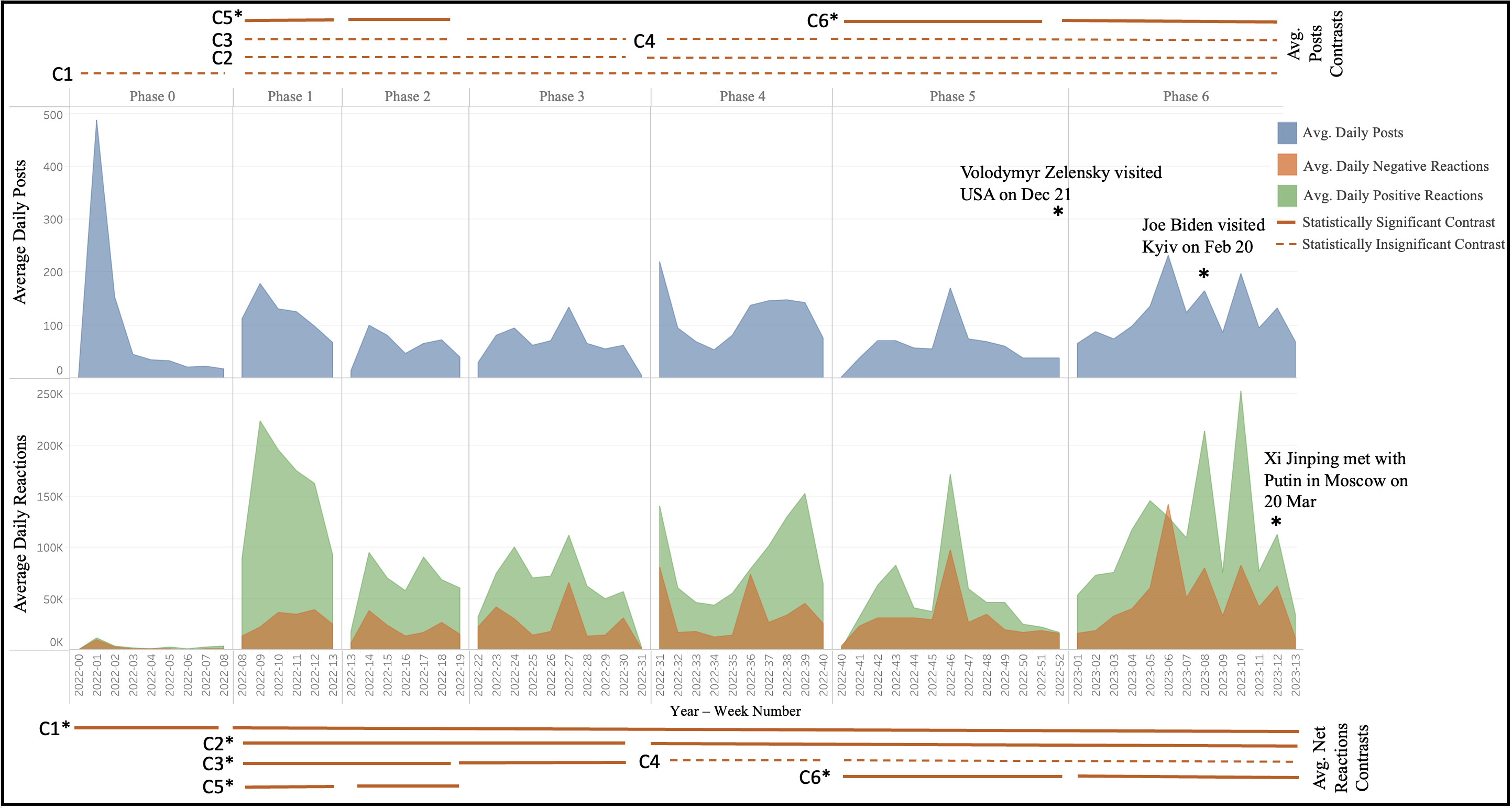}  
    \caption{International Politics theme across the phases. We captured the (i) evolution of average weekly posts (blue area chart), (ii) evolution of positive and negative user engagement (green and orange area chart), (iii) statistical significance of contrast weights for post counts and net reactions (dotted bars/single line bars)}
    \label{Fig:International_Politics}
\end{figure}

Our next set of analyses demonstrates the capability for repositioning within Anti-Kremlin channels (see Figures \ref{Fig:Combat} and \ref{Fig:International_Politics}). Here, we focus on the topics of \textit{“International politics”} and \textit{“Combat and Frontline updates”}, where the significant contrasts regarding combat are C3, C4, C5, and C6. Our point here is that the pattern of significant contrasts is different relative to ‘economic concerns’ and between the two topics. Although C2, C5, and C6 are significant, the direction of C5 and C6 is different. In addition, C3 and C4 become significant. The increase of Combat and Frontline updates aligns with specific external events, such as the initial escalation (phases 1 and 2) after invasion (C5), de-escalation (Phases 3 and 4; C3) following Ukraine’s effective defense (Wulf, 2022), and a subsequent increase after  Russia's retaliatory response to Ukraine’s Kerch bridge bombing (after phase 4; C4 and C6) \citep{Murauskaite2023}. The trend in the Phases 5 and 6 documents increases (C6) and not decreases, particularly during and after significant offline events, such as Ukraine’s counter-offensive faced with Russian retaliation and Zelenskyy’s US visit, with apparently higher post volumes relative to economic concerns. The ebb and flow of combat update topics supports the claim that combat is being exploited in lieu of Russian economic issues (see Figure \ref{Fig:Combat}). User engagement generally follows suit with C3 and C4, with increasing user engagement relative to the pattern of interest in the economic topic in later phases. 

Figure \ref{Fig:International_Politics} shows the dynamic of international politics, a pervasive topic in Anti-Kremlin channels. Here, we see significant effects in C5 (as a decline), as well as an increase in C6, clearly aligned with specific events. Contrast 5, during the initial phase of the invasion, was characterized by extensive discussion on international involvement and sanctions on Russia. Phase 6 included important geopolitical events such as Ukrainian President Zelensky's visit to the U.S., and the subsequent influx of Western military aid, as well as U.S. President Biden’s visit to Ukraine and Chinese President Xi Jinping's visit to Russia. Relative to combat, C3 and C4 are not significant.  Engagement is similarly U-shaped with significant C3 and C6. 

\begin{figure}[h!]
    \centering
    \includegraphics[width=\linewidth]{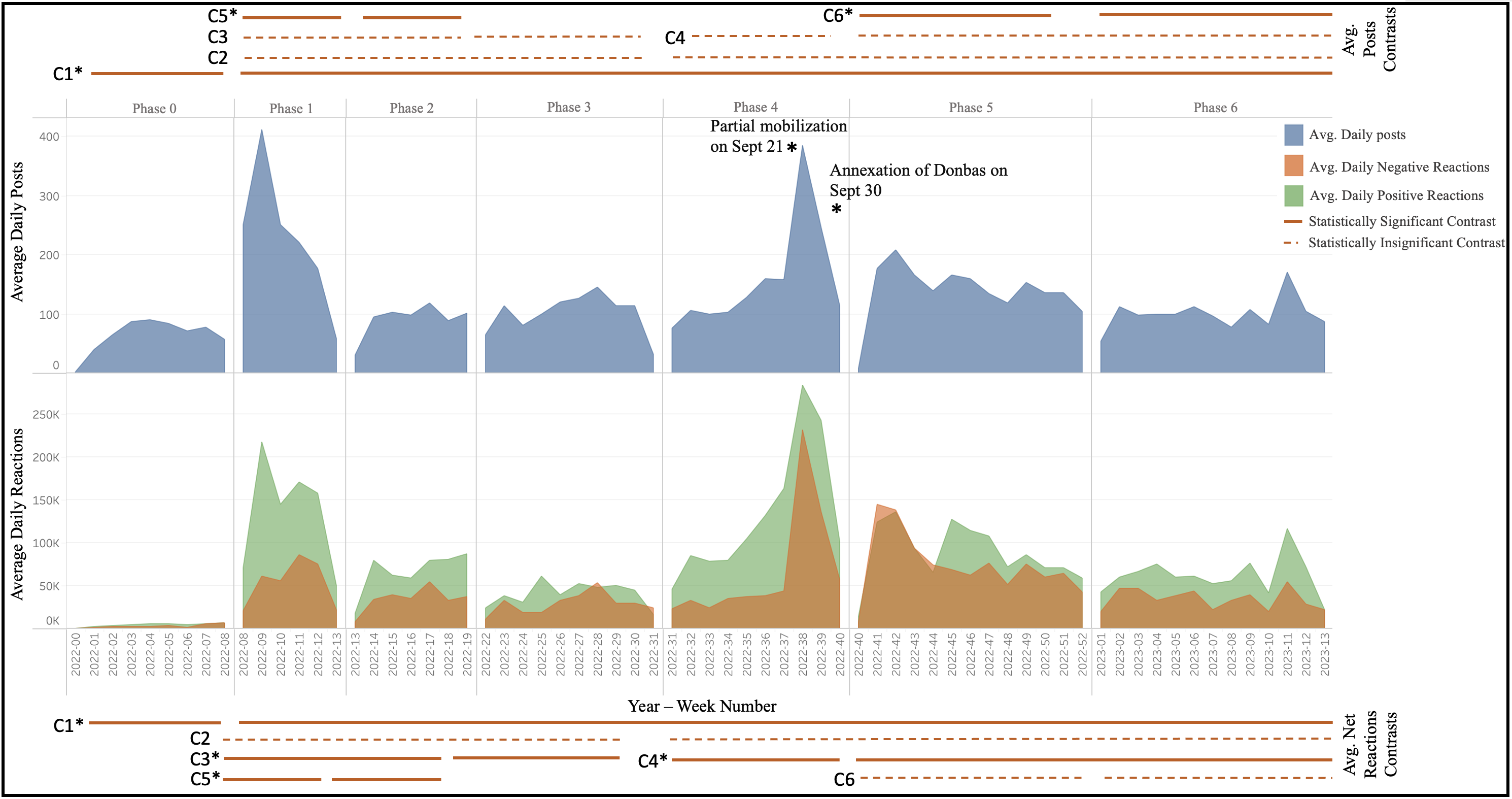}  
    \caption{Temporal Dynamics of Russian Domestic Affairs and Viewer Reactions Across Phases. Increased negative reactions to "Russian Domestic Affairs" during significant events (e.g., the annexation of Donbas and partial mobilization) indicate potentially elevated public discontent. Post volumes and user engagement trends, especially during key political events, reflect the Russian opposition's adaptive capability of influencing public sentiment. The blue line depicts weekly average post volumes, while the green and orange areas show positive and negative reactions, respectively. Blue and orange bars indicate statistical significance from contrast tests for post volumes and viewer reactions, respectively.}
    \label{Fig:Russian_Dom_Aff}
\end{figure}

\subsubsection{Finding 3: The Kremlin does not Control the Anti-Kremlin Information Environment}\label{subsubsec2}

Figure \ref{Fig:Russian_Dom_Aff} illustrates the temporal dynamics of the \textit{“Russian domestic affairs”} theme.  C5 and C6 document a zig-zag pattern, unlike the functions we saw in combat and international politics. Not surprisingly, C1 is significant, consistent with the economic topic in Figure \ref{Fig:Economy}. User engagement follows the post pattern with significant contrasts 3, 4, and 5 (see also Table \ref{Table 3}). However, the striking finding here is the substantial increase in viewer negative reactions, especially during critical events with intense political relevance (C4), such as partial mobilization in Russia on September 21, 2022 (week 39, Phase 4), indicating elevated public unrest. That pattern is also associated with the announcement of the annexation of Donbas on September 30, 2022 (week 40, at the end of Phase 4). \bigskip

In contrast to heightened viewer reactions in Phase 4, Phase 5 and 6 reactions declined (C4) yet remained mostly negative under this theme. User engagement and the sentiments expressed in discussions on specific subcategories, such as \textit{“opposition crackdown”}, \textit{“mobilization”}, and \textit{“referendum”} under \textit{Russian Domestic Affairs}, indicate a potentially growing public discontent (see Table \ref{Table 1}). More specifically, user engagement with these discussions indicates strategies being developed to avoid or counter repercussions upon certain anti-Kremlin actions, such as discrediting Russian armed forces (Appendix Figure \ref{Appendix: Figure B3}), evading mobilization (see Appendix Figures \ref{Appendix: Figure B4}, \ref{Appendix: Figure B5}), and encouraging Russian soldiers to lay down their weapons (Appendix Figure \ref{Appendix: Figure B8}), which attracted positive viewer reactions. On the other hand, critiques of mobilization and referendum (Appendix Figures \ref{Appendix: Figure B6}, \ref{Appendix: Figure B7}) generated negative viewer reactions, suggesting user support for resistance efforts. The pattern of posts and negative viewer reaction regarding Russian domestic affairs, in combination with the reaction to international affairs, is not favorable to the Kremlin. 

\section{Discussion}
\label{sec3}

Our study investigates online communications during the ongoing Russia-Ukraine conflict, focusing on 114 Anti-Kremlin channels with over 1 million posts and corresponding viewer reactions. Our results argue for the exploitation of attention-garnering breach using changes in economic affairs, combat and frontline and international politics as an adaptive communication strategy synchronized with external events. The argument benefits from naturalistic longitudinal big data involving the same agent battling for public attention.  Future work will address the inevitable limitations of a first study of anti-Kremlin online communications during the Russian invasion of Ukraine. \bigskip

Two analyses demonstrated the synchronization of content with external events.  First, the channels attempted to exploit a predicted downturn in the Russian economy \citep{Sonnenfeld2022}. The effort was abandoned, consistent with the observation that Western sanctions did not, in fact, affect the Russian economy in the short term \citep{Egorov2023}. Second, drawing on narrative theory, we demonstrated a change in topic prominence coincident with world events regarding culturally relevant exceptional, non-routine combat and international affairs. By adapting their content synergistically, opposition channels effectively prioritized certain critical issues at each phase of the conflict, reflecting the shifting dynamics on the ground to maintain a stream of salient, breach-worthy posts. Adaptive messaging in social media is not new; we see it particularly in the form of culturally-sensitive marketing \citep{Chandra2022}. What is notable here is being responsive to sequential events, consistent with a growing concern for the dynamic environment in political communication \citep{Perloff2021}. While our focus here is on political communication, our emphasis on the dynamics of communication strategies resulting from the speed of online distribution applies to any domain constantly subject to external perturbations. Tourism messaging, for example, must respond to breach events such as natural disasters and pandemics \citep{Duro2021,Filimonau2020}. \bigskip

We capped the analyses revealing adaptation with a demonstration that the communication strategy was effective regarding the Anti-Kremlin mission. In parallel, by engaging viewers on the very topics (e.g., “suppression of free speech”, “mobilization”, and “referendum”) that the Kremlin had been deliberately avoiding and/or preventing other channels from reporting \citep{Baysha2024,Troianovski2023,Troianovski2024}, we show that the Kremlin has lost control of the narrative. An event such as the annexation of Donbas, which should have been viewed as a success from the Kremlin’s perspective, elicited a decidedly negative response. We saw similar audience approval when Russian banks were removed from the SWIFT banking system. These findings suggest that Anti-Kremlin channels are actively shaping the online discourse around the ongoing conflict, focusing on the breach topics that the Kremlin and its strictly controlled mainstream media have been intentionally avoiding or censoring. These efforts comprise counter-speech to challenge Kremlin narratives, demonstrating the opposition’s resilience amidst political turmoil and a sophisticated understanding and application of propaganda strategies. Subsequent public and local governmental support for Prigozhin's unsuccessful coup attempt against the Kremlin reinforces our impression of a potential rise in public discontent with the regime. The Kremlin’s attempts to suppress these narratives with crackdowns on opposition figures and journalists seem to have limited influence and reach within the opposition channels. \bigskip

To make these points, we conducted a longitudinal analysis of big data, spanning seven phases of conflict. Unlike the preponderance of social media content analyses that focus on citizen journalism and the characterization of public opinion \citep{Kloo2024}, we focused on the intentional posting behavior of an organized social movement. Data-driven computational topic analysis allowed us to characterize this content. The public response here serves as feedback on posting behavior in a dynamic that measures public interest and continued engagement. We backed our assertions with statistical analysis, opting for a categorical analysis of phases, defined by an existing source, coincident with critical external events. In comparison to classical regression analysis, our a–priori contrast-based approach captured complex functions of change over time, and allowed us to identify differences across different content measures using the same basic framework. We also reduced vulnerability to multiple paired comparisons. The identified statistically significant thematic patterns across phases suggest that Anti-Kremlin propaganda content was not random, signifying synergistic efforts in relation to emerging offline events. \bigskip

Our study, while comprehensive, is subject to some limitations that suggest avenues for future research. This study did not account for the influence of bots and trolls. This is not a problem for the Anti-Kremlin sources, but bots or trolls could have influenced measures of user engagement. Furthermore, while our analysis withstood a conservative low-power approach, a more granular analysis could reduce measurement error and better capture the peaks and variations within phases, offering a finer-grained picture of the adaptive communication strategies employed by these channels. On the other hand, such an analysis would also lead to more tests and spurious findings enabled by higher power. We make no claims that the particular pattern of significant contrasts applies to a different set of event-based phases, but rather that the general approach facilitates comparison between measures. Finally, although we have tended to attribute intentionality to the absence of randomness in our statistical tests, a simpler mechanism is viable. Anti-Kremlin posters may not be following an explicit breach-maximization rule, but rather simply reacting to, and posting, breach content they find of interest, and like-minded viewers respond. What looks like rule-following may actually result from an emergent implicit process characteristic of dynamical systems. \bigskip

Future work will address several important aspects that were beyond the scope of this study. First, a synchronized analysis involving Pro-Kremlin channels could enhance our understanding of the broader online discourse dynamic. Second, incorporating network analysis to explore how these channels interact, including the forwarding of information and the network of users across both Anti-Kremlin and Pro-Kremlin channels, would provide key insights into the dissemination and influence patterns. Additionally, extending our analysis beyond the seven phases up to March 2023 to include subsequent phases would help capture the ongoing evolution of Anti-Kremlin rhetoric and its responses to new developments.


\section{Methods}
\label{sec4}

\begin{figure}[h!]
    \centering
    \includegraphics[width=\linewidth]{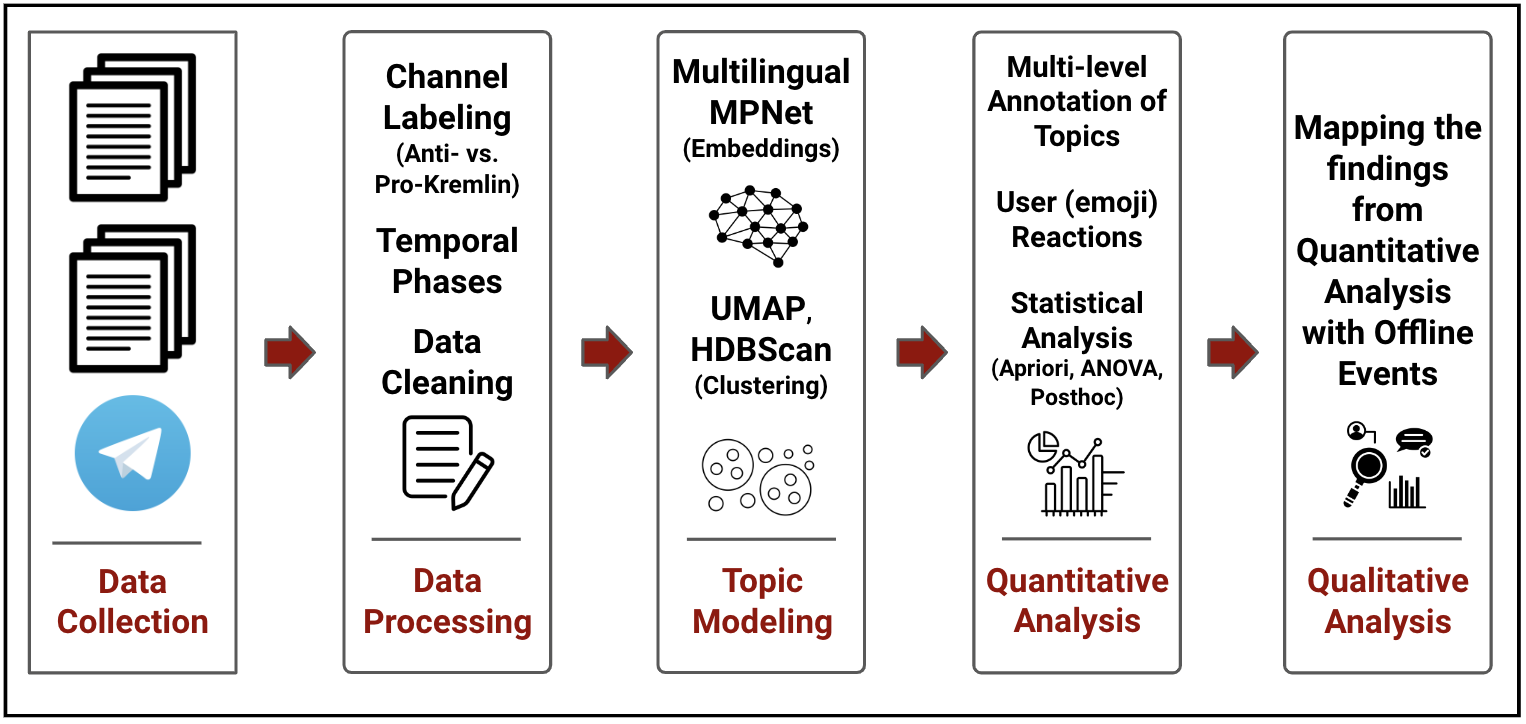}  
    \caption{Our approach for analyzing political content and viewer reactions via emoji related to Russian politics from Telegram channels.} 
    \label{fig: Architecture}
\end{figure}

Figure \ref{fig: Architecture} provides an overview of our analysis approach. We qualitatively labeled the Telegram channels as Anti-Kremlin and Pro-Kremlin. Based on the phases of the conflict, according to the timeline by 
\citet{Murauskaite2023}, we placed the data in their respective temporal phases and processed them using natural language processing methods. We utilized the multilingual MPNet model to generate embeddings of the Russian language and create topical clusters for each phase, which we qualitatively annotated to reveal higher-level topical categories and viewer reactions via emoji. Then, we performed statistical analyses of the quantitative patterns observed across the phases within an ANOVA framework. Lastly, we performed a qualitative analysis of the quantitative findings in the context of offline events in their respective phase of the conflict. 

\subsection{Anti-Kremlin Dataset}\label{subsec2}

We first collected data from the most prominent Russian Telegram channels. Specifically, we identified and selected the 614 Russian political channels with the largest subscriber base, with at least 10,000 subscribers from the TGStat platform \citep{TGStat2024}, which provides a catalog of Telegram channels categorized by country, language, and themes. A native Russian-speaking coder manually labeled channels as \textit{Pro-Kremlin}, \textit{Anti-Kremlin}, or \textit{neutral} based on the following criteria: The channels that regularly post state-sponsored Kremlin propaganda were labeled as \textit{Pro-Kremlin}, while those criticizing the Russian President Putin and the Kremlin were labeled as \textit{Anti-Kremlin}. Channels that primarily shared news without discernable bias were marked as neutral. A non-Russian-speaking coder subsequently verified this first-level annotation using Telegram’s English translation feature. Our Russian-speaking political science domain expert resolved the conflicts and performed final validation through the verification of randomly selected samples for robustness and reliability. A Kappa score of 0.89 between the domain expert and coder indicated substantial agreement with the expert's annotations. We obtained 402 Pro-Kremlin and 114 Anti-Kremlin channels with data points of 3,571,787 and 1,354,084 posts, respectively, along with 99 neutral channels. \bigskip

We focused on the 114 Anti-Kremlin channels, analyzing the patterns in information disseminated before and after the invasion with respect to offline events. We narrowed the dataset to 354,819 posts from January 1, 2022, to March 31, 2023, including: (i) text content of each post, (ii) associated (images, videos), (iii) date and time, (iv) view counts, (v) forward counts, (vi) original or forwarded, (vii) source channel if forwarded, (viii) emoji viewer reactions and their counts, (ix) user replies with similar meta-data such as text, date and time, and reactions. 

\subsection{Temporal Phases}\label{subsec2}
\begin{figure}[h!]
    \centering
    \includegraphics[width=\linewidth]{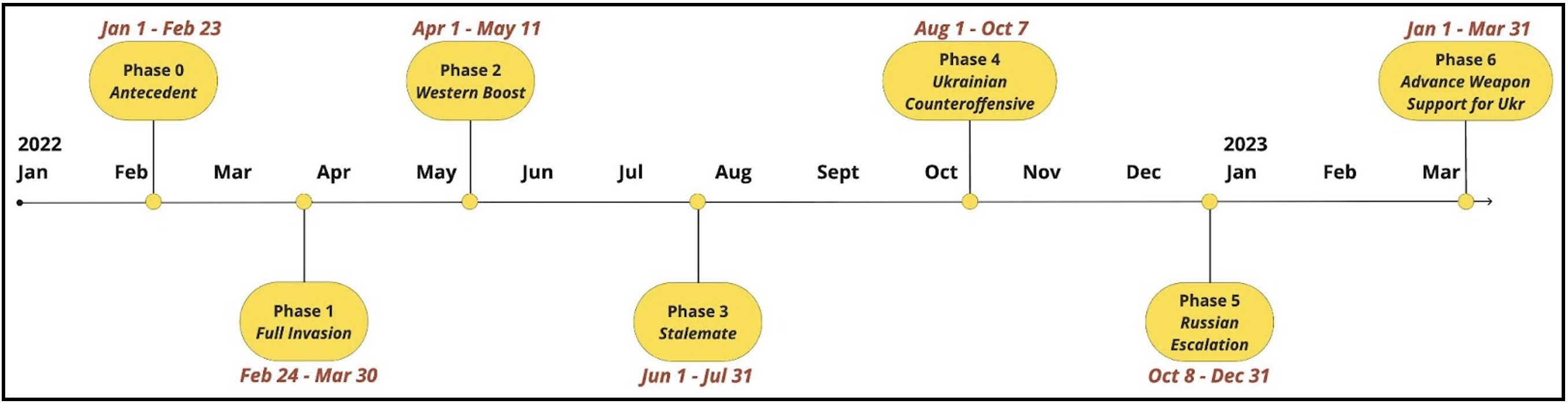}  
    \caption{Temporal phases of the war, phase 0 is pre-invasion, and phase 1 to phase 6 are post-invasion phases. The U.S. Arms Transfers to Ukraine study includes phases 0 to 5. Considering the critical events of late 2022 and early 2023, we further include phase 6 from January 1 to March 31, 2023.}    
    \label{fig: phases}
\end{figure}

We examined the impact of offline events on online discourse within Anti-Kremlin Telegram channels and viewer reactions to these posts related to specific offline events during the conflict. We followed the temporal phases defined by The National Consortium for the Study of Terrorism and Responses to Terrorism (START) at the University of Maryland \citep{Murauskaite2023}. Using this framework, our study segmented the narrative evolution across the seven phases, comprising 22,511 data points for the pre-invasion phase and 332,308 for six post-invasion phases, as shown in Figure \ref{fig: phases}. As these phases are based on offline events; phases can begin and end mid-week, and there may be brief time gaps between phases. Descriptive statistics of the dataset across the phases are provided in Appendix Table \ref{Appendix: Table A5}. 

\subsection{Topical Analysis}\label{subsec2}
We grouped our dataset into fine-grained topical clusters to identify recurring themes, evolving narratives, and emerging trends within Anti-Kremlin channels throughout the conflict. We employed several language processing steps, which included removing punctuation, hyperlinks, numbers, special characters, and Russian stopwords \citep{zafarani2014social,kursuncu2019predictive}. In addition, lemmatization was employed to address word inflections in the Russian language, using the MyStem package \citep{Pymystem2018} for morphological analysis. This process standardized variations of the word “Russian” (e.g., “российской”, “российских”, “российские”) to their lexical root. For topical modeling, we employed the BERTopic method (Grootendorst 2022) for each of the seven phases of the conflict, utilizing the multilingual version of the MPNet \citep{Song2020} to generate 768-dimensional contextual word embeddings of Russian text. The posts had a mean token length of 445, compared to the MPNet’s token length 512. We reduced the embedding dimensions to five using UMAP \citep{McInnes2018}, tuning the parameters to maintain the integrity of the information. Then, we finalized the 15 nearest neighbors using the Euclidean distance metric. We used HDBSCAN to cluster these embeddings, where the top 10 keywords per cluster were identified as representing topical information using TF-IDF. Anticipating different topic prevalences by phase, we trained seven different topic models ranging from 35 to 200 topics to identify the most optimally representative model for each of the conflict's seven phases (six post-invasion and one pre-invasion). We assessed the quality of topics generated for each phase of the conflict through a coherence analysis of our topic models \citep{Syed2017,kursuncu2019modeling,gaur2018let}. Topic coherence is crucial as higher coherence scores indicate topics that are more semantically interpretable and meaningful to humans. We utilized CV coherence scores and elbow technique to determine the most representative topic model for each phase. The optimal model for each phase was chosen based on these coherence scores, with the number of topics selected ranging from 50 to 200 (see Appendix Figure \ref{Appendix: Figure B2}). \bigskip

Further, we conducted a qualitative analysis of these topics using the Gioia method \citep{Gioia2013}. First, the top 10 keywords per topic were translated into English using the Google Translator API \citep{GoogleTrans2020}, allowing analysis in both Russian and English versions. Two researchers from our team, including one native Russian speaker, annotated these topics according to \textit{thematic first-order} categories and \textit{second-order} (sub) categories for each phase. An independent political science scholar, also a native Russian speaker, reviewed this annotation for accuracy and validation. To align the online communications with real-world events, we mapped offline events documented by the Institute for Study of War (ISW) \citep{Clark2022} (see Appendix \ref{Appendix C}) onto the timeline of conflict (i.e., seven temporal phases), assessing the relevance of the topical categories to the offline events. We corroborated these offline events with both academic and traditional media sources (see Appendix \ref{Appendix C}), including the “Russian War and Invasion of Ukraine” timeline by Purdue University \citep{Tsymbaliuk2024}, as well as coverage from “CNN” \citep{CNN2023}, and “Euronews” \citep{Askew2023}.

\subsection{Dependent Measures}\label{subsec2}
We focus on two types of dependent measures to assess the dynamics of the online Anti-Kremlin narrative: topic post volume and viewer reactions via emoji. These measures provide insights into the prevalent themes, emerging topical trends, and user engagement and sentiment. \bigskip

Post Volume is defined as the number of posts created within a topical category in a given time period, serving as a key measure to gauge the interest level in specific topics. This study quantifies post volume using Equation \ref{eq: ANOVA}, which computes the daily average number of posts within a week per topic. 

\begin{equation}
\forall c \in C, \, \mu_i^c = \frac{1}{k} \sum_{j=1}^{k} n_j^c \sim \mathcal{C}(phase)
\label{eq: ANOVA}
\end{equation}

\textit{C} denotes the set of all topics, \textit{c} represents individual topics in \textit{C}, $\kappa$ is the number of days in a given week, \( n_j^c \) is the number of posts for a given day j, and \( \mu_i^c \) is the mean post volume for a given week \textit{i} in a phase. As phases can begin and end mid-week, we used $\kappa$ as the number of days in a week. In Figure 1, min-max normalization was applied to the post volumes across the weeks throughout the phases, to facilitate the qualitative identification of relative changes independent of fluctuations in the overall volume. \bigskip

Viewer Reactions on Telegram are primarily expressed through emoji. We categorized these viewer reactions into positive and negative sentiments based on the type of emoji, following EmojiNet \citep{Wijeratne2017}. Positive reactions included emoji, such as “\includegraphics[height=1em]{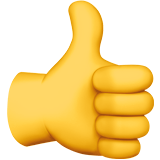}” (thumbs up), “\includegraphics[height=1em]{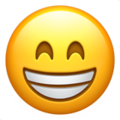}” (grinning face), “\includegraphics[height=1em]{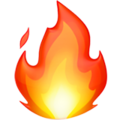}” (fire), “\includegraphics[height=1em]{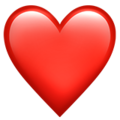}” (heart), “\includegraphics[height=1em]{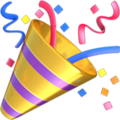}” (party popper), “\includegraphics[height=1em]{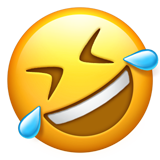}” (rolling on the floor laughing), and “\includegraphics[height=1em]{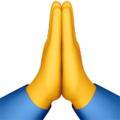}” (folded hands). On the other hand, emoji, such as “\includegraphics[height=1em]{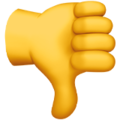}” (thumbs down), “\includegraphics[height=1em]{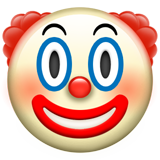}” (clown face), “\includegraphics[height=1em]{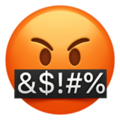}” (face with symbols on the mouth), “\includegraphics[height=1em]{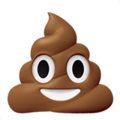}” (pile of poo), “\includegraphics[height=1em]{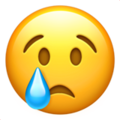}” (crying face), and “\includegraphics[height=1em]{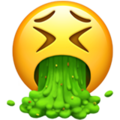}” (vomiting) were categorized as negative reactions. We calculated the net sentiment for each post by subtracting the count of negative emoji from positive emoji and averaged these values on an average daily measure for each week. Then, we applied equation \ref{eq: ANOVA}, where \( n_j^c \) is the number of viewer reactions for a given day j.

\subsection{Contrast Tests on Dependent Measures}\label{subsec2}

\begin{table}[h!]
\centering
\begin{tabular}{lccccccc}
\toprule
\textbf{Contrasts} & \multicolumn{7}{c}{\textbf{Phases}} \\ 
\cmidrule(lr){2-8}
 & \textbf{Phase 0} & \textbf{Phase 1} & \textbf{Phase 2} & \textbf{Phase 3} & \textbf{Phase 4} & \textbf{Phase 5} & \textbf{Phase 6} \\ 
\midrule
\textbf{Contrast 1} & -6 & 1 & 1 & 1 & 1 & 1 & 1 \\ 
\midrule
\textbf{Contrast 2} & 0 & 1 & 1 & 1 & -1 & -1 & -1 \\ 
\midrule
\textbf{Contrast 3} & 0 & -1 & -1 & 2 & 0 & 0 & 0 \\ 
\midrule
\textbf{Contrast 4} & 0 & 0 & 0 & 0 & 2 & -1 & -1 \\ 
\midrule
\textbf{Contrast 5} & 0 & 1 & 1 & 0 & 0 & 0 & 0 \\ 
\midrule
\textbf{Contrast 6} & 0 & 0 & 0 & 0 & 0 & 1 & -1 \\ 
\bottomrule
\end{tabular}
\caption{Orthogonal contrasts to compare the specific scenarios. Contrast 1 compares the pre-invasion phase with the post-invasion phase; Contrast 2 compares phases 1, 2, and 3 vs. phases 4, 5, and 6; Contrast 3 compares phases 1, 2 vs. phase 3; Contrast 4 compares phase 4 vs. phases 5 and 6; Contrast 5 compares phase 1 vs. phase 2; and Contrast 6 compares phase 5 vs. phase 6.}
\label{Table 4}
\end{table}

Following the annotation of topics, we conducted a statistical analysis using a one-way Analysis of Variance (ANOVA) to examine variations in post volumes between different event-based phases of the conflict. This involved verifying ANOVA’s assumptions concerning the homogeneity of variance and outlier removal. Then, drawing on the one-way ANOVA framework, a common set of apriori contrasts and post-hoc tests of post volume by topic over phases were conducted, using a two-tailed t-test with pooled error terms based on variability between the weeks associated with a phase rather than the typical regression based high-power daily fluctuations. This approach constrains the number of statistical tests and provides a common framework for comparing changes between topical categories liberated from the typical trend analysis weights. As shown in Table \ref{Table 4}, the contrasts defined in our analysis include examining: (i) the pre-invasion phase (Phase 0) and all subsequent phases (Contrast 1). (ii) differences between the timeline's first and second halves (Contrast 2). (iii) differences between the beginning and the final phases of each half (Contrasts 5 and 6), and (iv) the changes within the two remaining intermediate phases in each half (Contrast 3 and 4). 

The additional qualitative analysis maps the identified prominent themes with the corresponding offline events across different phases.


\bibliography{sn-bibliography}

\clearpage

\begin{appendices}

\section{Tables}\label{secA1}

\begin{table}[h!]
\centering
\begin{tabular}{p{2.5cm}p{1.5cm}p{1.1cm}p{1cm}p{1.4cm}p{1.2cm}p{1.4cm}}
\toprule 
\textbf{Categories} & \textbf{Factor Variable \newline (per week)} & \textbf{Sum of Squares} & \textbf{ dof} & \textbf{Mean Square} & \textbf{F-ratio} & \textbf{P-value} \\ 
\midrule
\multirow{2}{10em}{Economy} & \centering posts & 140352.971 & \centering 6 & 23392.162 &  30.668 &  2.06E-16* \\ \cmidrule(lr){2-7}
 & \centering reactions &  7.95E+10 & \centering 6 &  1.33E+10 &  34.202 &  4.991e-11* \\ 
\midrule
\multirow{2}{5em}{International Politics} & \centering posts &  54099.093 & \centering 6 &  9016.516 &  1.743 &  0.127 \\ \cmidrule(lr){2-7}
 & \centering reactions &  9.33E+10 & \centering 6 &  1.56E+10 &  15.974 &  8.33E-11* \\ 
\midrule
\multirow{2}{8em}{Combat and Frontline Updates} & \centering posts &  308127.707 & \centering 5 &  61625.541 &  66.008 &  4.53E-21* \\ \cmidrule(lr){2-7}
 & \centering reactions &  1.22E+11 & \centering 5 &  2.44E+10 &  12.142 &  9.06E-08* \\ 
\midrule
\multirow{2}{10em}{ Russian Domestic Affairs} & \centering posts &  185051.756 & \centering 6 &  30841.959 &  5.954 &  0.000065* \\ \cmidrule(lr){2-7}
 & \centering reactions &  5.38E+10 & \centering 6 &  8.96E+09 &  15.394 &  1.58E-10* \\ 
\midrule
\multirow{2}{10em}{ Ukrainian Domestic Affairs} & \centering posts &  447742.278 & \centering 6 &  74623.713 &  24.903 &  1.72E-14* \\ \cmidrule(lr){2-7}
 & \centering reactions &  1.18E+11 & \centering 6 &  1.97E+10 &  2.59 &  0.027* \\ 
\bottomrule
\end{tabular}

\caption{Summary of ANOVA results for categories: \textit{Russian Domestic Affairs, Economy, Combat and frontline updates, Ukrainian domestic affairs, and International Politics.} Factor variables, post volumes and viewer reactions, are reported as average daily per week. SoSq: Sum of Squares, dof: Degree of Freedom.}
\label{Appendix: Table A1}
\end{table}

\newgeometry{margin=3cm,footskip=3em,}
\begin{table}[h!]
\centering
\begin{tabular}{p{1.8cm}p{1.8cm}p{9.5cm}}
\toprule
\textbf{Categories} & \textbf{Subcategories} & \textbf{Keywords \footnotesize(Original and English translation)} \\ 
\midrule
\multirow{3}{6em}{\parbox{1\linewidth}{\vspace{2.5cm}Russian Domestic affairs}} 
    & Opposition Crackdown / Suppression of free speech 
    & \textbf{In Russian:} суд, дело, обвинение, фейк, похищение, активист, задерживать, арестовать, пытка, осужденный, голодовка, антивоенный, митинг, иностранный, агент \newline
    \textbf{English translation:} court, case, accusation, fake, kidnapping, activist, detain, arrest, torture, convict, hunger strike, antiwar, rally, foreign, agent \newline
    \textbf{Mentioned personalities:} Alexei Navalny, Abubakar Yangulbaev, Daria Serenko, Vladimir Kara-Murza, Dmitry Kolker, Andrey Pivovarov, Ivan Safronov, Ilya Yashin, Rita Flores \\ \cmidrule(lr){2-3}
    & Mobilization 
    & \textbf{In Russian:} содержание, особый, приказывать, переводить, мобилизация, частичный, военный, протест, задерживать, митинг, полиция, война, военкомат, уход, отставка, министр \newline
    \textbf{English translation:} containment, special, order, transfer, mobilization, partial, military, protest, detain, protest, rally, police, war, military registration and enlistment office, PMC, resignation, minister \newline
    \textbf{Mentioned cities:} Dagestan, Buryatia, Bashkortostan, Mordovia, Belgorod, Perm \\ \cmidrule(lr){2-3}
    & Referendum 
    & \textbf{In Russian:} референдум, губернатор, выборы, отставка, голосование, присоединение, муниципалитет, партия, сентябрь, территория, оккупировать \newline
    \textbf{English translation:} referendum, governor, elections, resignation, vote, annexation, municipal, party, september, territory, occupy \\ 
\midrule
\multirow{3}{6em}{\parbox{1\linewidth}{\vspace{2.2cm}Economy}} 
    & Currency 
        & \textbf{In Russian:} рубль, доллар, евро, падение, фондовый, торговать, банк, дефолт, московский, биржа, исторический, свифт, отключение, инфляция, ввп, кризис, дефолт, инфляция \newline
        \textbf{English translation:} ruble, dollar, euro, fall, stock, bargaining, bank, default, moscow exchange, historical, swift, shutdown, inflation, GDP, a crisis, default, inflation \\ \cmidrule(lr){2-3}
    & Gas/Oil 
        & \textbf{In Russian:} газ, поставка, цена, превышать, потолок, газопровод, поток, остановить, Contract, оплата, нефть, импорт, эмбарго, экспорт, утечка \newline
        \textbf{English translation:} gas, supply, price, exceed, ceiling, pipeline, flow, stop, contract, payment, oil, import, embargo, export, leak \newline
        \textbf{Mentioned countries:} Germany, Europe, USA, Poland, Saudi Arabia \newline
        \textbf{Mentioned companies:} GAZPROM, NK ROSNEFT \\ \cmidrule(lr){2-3}
    & Food Export 
        & \textbf{In Russian:} зерно, пшеница, продовольственный, порт, экспорт, оон, вывозить, глобальный, судно, турция, одесса, коридор, черноморский, сделка, продление \newline
        \textbf{English translation:} grain, wheat, food, port, export, UN, export, global, ship, Turkey, Odessa, corridor, Black Sea, deal, extension \\ 
\midrule
\multirow{1}{6em}{International politics} 
        & - & \textbf{Mentioned countries/topics:} Kazakhstan, Georgia Europe Visa, China, USA, Iran, Belarus, Turkey, Azerbaijan, Armenia, Israel, Moldova \\ 
\midrule
\multirow{2}{6em}{\parbox{1\linewidth}{\vspace{0.7cm}Ukrainian Domestic Affairs}} 
    & Electricity Shutdown 
        & \textbf{In Russian:} мост, крымский, исключение, отключение, электричество, взрыв, электроэнергия, expl, электроснабжение, крым, энергорго \newline
        \textbf{English translation:} bridge, Crimean, included, shutdown, electricity, explosion, electricity, fire, light, Ukrenergo \\ \cmidrule(lr){2-3}
    & Refugee Mobilization 
        & \textbf{In Russian:} беженец, обмен, черншук, украинский, пленник, покинуть, миллион, гражданство, паспорт, беженец, виза, указ, оон \newline
        \textbf{English translation:} refugee, exchange, Yershchuk, Ukrainian, captive, leave, million, citizenship, passport, refugee, visa, decree, UN \\ 
\midrule
\multirow{1}{6em}{\parbox{1\linewidth}{\vspace{0.3cm}Combat and Frontline updates}} 
        & - & \textbf{In Russian:} российский, военный, ракета, танк, оккупанты, вооруженные, армия, украина, враг, дрон, атаковать, самолет, обстрел, снаряд, жилой, война, удар, ракета \newline
        \textbf{English translation:} Russian, military, rocket, tank, occupants, Armed Forces of Ukraine, enemy, drone, attack, airplane, shoot, shelling, residential, war, hit, missile, perish, booth, himars, eagle, Bucha \\ 
\midrule
\end{tabular}%

\caption{Keywords in original and English translation across various categories and subcategories related to the conflict.}
\label{Appendix: Table A2}
\end{table}
\restoregeometry

\begin{table}[h!]
\centering
\begin{tabular}{p{0.6cm}p{1.5cm}p{7.6cm}p{1.5cm}}
\toprule
\textbf{Phase} & \textbf{Online Themes} & \textbf{Key Offline Events} & \textbf{Viewer Reactions} \\ \midrule
1 & Economy & Imposed sanctions and its anticipated impact on the Russian economy (\textit{Event A}) & Pos:88.5\%, Neg:11.5\% \\ \midrule
2 & War and Combat updates & Increase in intensity of war after Western boost to Ukrainian counteroffensive (\textit{Event B}) & Pos:65.5\%, Neg:34.5\% \\ \midrule
3 & Economy & Russia defaulted on its external sovereign bonds for the first time in a century and billion dollars in aid to Ukraine (\textit{Event C}) & Pos:66.1\%, Neg:33.8\% \\ \midrule
4 & Russian domestic affairs & Declare partial mobilization in the Russian Federation/Russian annexation of occupied Ukrainian territory including Donbas (\textit{Event D}) & Pos:66.7\%, Neg:33.3\% \\ \midrule
5 & Ukrainian domestic affairs & Russian forces launched a massive wave of strikes against critical Ukrainian infrastructure in retaliation over the destruction of Kerch Strait bridge (\textit{Event E}) & Pos:63.5\%, Neg:36.5\% \\ \midrule
\multirow{2}{*}{\parbox{1\linewidth}{\vspace{1cm}6}} & International politics & Joe Biden visited Kyiv, Xi Jinping visited Russia, Iranian Foreign Minister met Russian Foreign Minister, and \$350 million of security assistance to Ukraine by the US (\textit{Event F}) & Pos:68.6\%, Neg:31.4\% \\ \cmidrule(lr){2-4}
 & War and Combat updates & Again, increase in the intensity of war in relation to Event F. (\textit{Event G}) & Pos:67.4\%, Neg:32.6\% \\ 
\bottomrule
\end{tabular}
\caption{The mapping between prominent online themes, corresponding key offline events, and positive or negative viewer reactions across conflict phases. Economic sanctions and related discussions in Phase 1 seemed to have garnered positive reactions, suggesting potential support among the users. Increased military actions and political developments, especially upon significant Western support for Ukraine, dominated the discussions, with users leaning more positively. The online discussions within anti-Kremlin channels were responsive and dynamically changing, adapting to the ongoing offline, real-world events, with users contributing to public sentiment. See Figure 4 for the timeline's confluence of these online and offline signals.}
\label{Appendix: Table A3}
\end{table}

\begin{table}[h!]
\centering
\begin{tabular}{p{2.1cm}p{1.3cm}ccccccc}
\toprule
\textbf{Categories} & \footnotesize\textbf{Factor Variable \tiny(\# weeks)} & \textbf{P. 0} & \textbf{P. 1} & \textbf{P. 2} & \textbf{P. 3} & \textbf{P. 4} & \textbf{P. 5} & \textbf{P. 6} \\ \midrule
\multirow{2}{*}{ Economy} & posts &  8 &  6 &  7 &  9 &  10 &  13 &  13 \\ \cmidrule(lr){2-9}
 &  reactions &  8 &  6 &  7 &  9 &  10 &  13 &  13 \\ \midrule
\multirow{2}{8em}{ International Politics} &  posts &  8 &  6 &  7 &  9 &  10 &  13 &  13 \\ \cmidrule(lr){2-9}
 &  reactions &  8 &  6 &  7 &  9 &  10 &  13 &  13 \\ \midrule
\multirow{2}{8em}{ Combat and Frontline Updates} &  posts &  - &  5 &  7 &  9 &  10 &  13 &  13 \\ \cmidrule(lr){2-9}
 &  reactions &  - &  5 &  7 &  9 &  10 &  13 &  13 \\ \midrule
\multirow{2}{8em}{ Russian Domestic Affairs} &  posts &  8 &  6 &  7 &  9 &  10 &  13 &  13 \\ \cmidrule(lr){2-9}
 &  reactions &  8 &  6 &  7 &  9 &  10 &  13 &  13 \\ \midrule
\multirow{2}{8em}{ Ukrainian Domestic Affairs} & posts &  8 &  6 &  7 &  9 &  10 &  13 &  13 \\ \cmidrule(lr){2-9}
 &  reactions &  8 &  6 &  7 &  9 &  10 &  13 &  13 \\ 
\bottomrule
\end{tabular}
\caption{Number of weeks covered in each phase for posts and reactions pertaining to various categories. P: Phase.}
\label{Appendix: Table A4}
\end{table}

\begin{table}[h!]
\centering
\begin{tabular}{p{4cm}ccccccc}
\toprule
 & \textbf{P. 0} & \textbf{P. 1} & \textbf{P. 2} & \textbf{P. 3} & \textbf{P. 4} & \textbf{P. 5} & \textbf{P. 6} \\ \midrule
\textbf{Number of weeks} &  9 &  6 &  7 &  9 &  10 &  13 &  14 \\ \midrule
\textbf{Number of Days} &  54 &  35 &  41 &  61 &  68 &  85 &  90 \\ \midrule
\textbf{mean post volume (avg. daily posts per week) / \# of weeks} &  429 &  1732 &  987 &  797 &  803 &  852 &  625 \\ \midrule
\textbf{Median post volume} &  353 &  1420 &  1001 &  790 &  799 &  803 &  638 \\ \midrule
\textbf{Standard Deviation in post volume} &  241 &  718 &  62 &  21 &  140 &  100 &  104 \\ \midrule
\textbf{Min avg post volume} &  136 &  1153 &  914 &  774 &  621 &  742 &  399 \\ \midrule
\textbf{Max avg post volume} &  966 &  2930 &  1073 &  838 &  1069 &  1023 &  779 \\ 
\bottomrule
\end{tabular}
\caption{Descriptive statistics of weekly post volume across different phases. P: Phase.}
\label{Appendix: Table A5}
\end{table}

\section{Figures}\label{secA2}

\begin{figure}[H] 
    \centering
    \includegraphics[width=\textwidth]{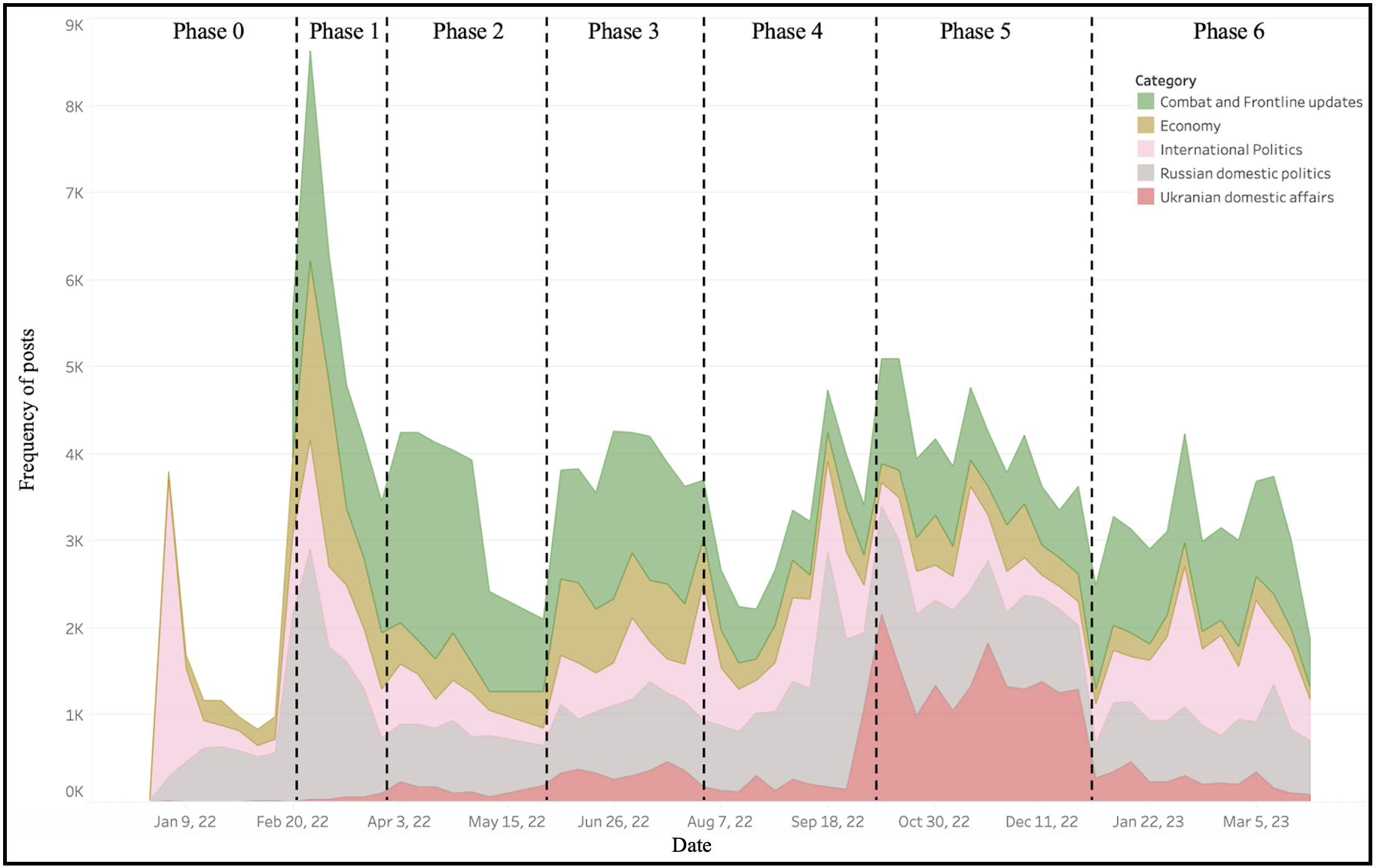}
    \caption{Absolute post volume Timeline (each category). The area chart represents the evolution of the top 5 categories over seven phases. Categories are represented as Combat and frontline updates, Economy, International Politics, Russian domestic affairs, Ukrainian domestic affairs.}
    \label{Appendix: Figure B1}
\end{figure}

\begin{figure}[H]
    \centering
    \includegraphics[width=\textwidth]{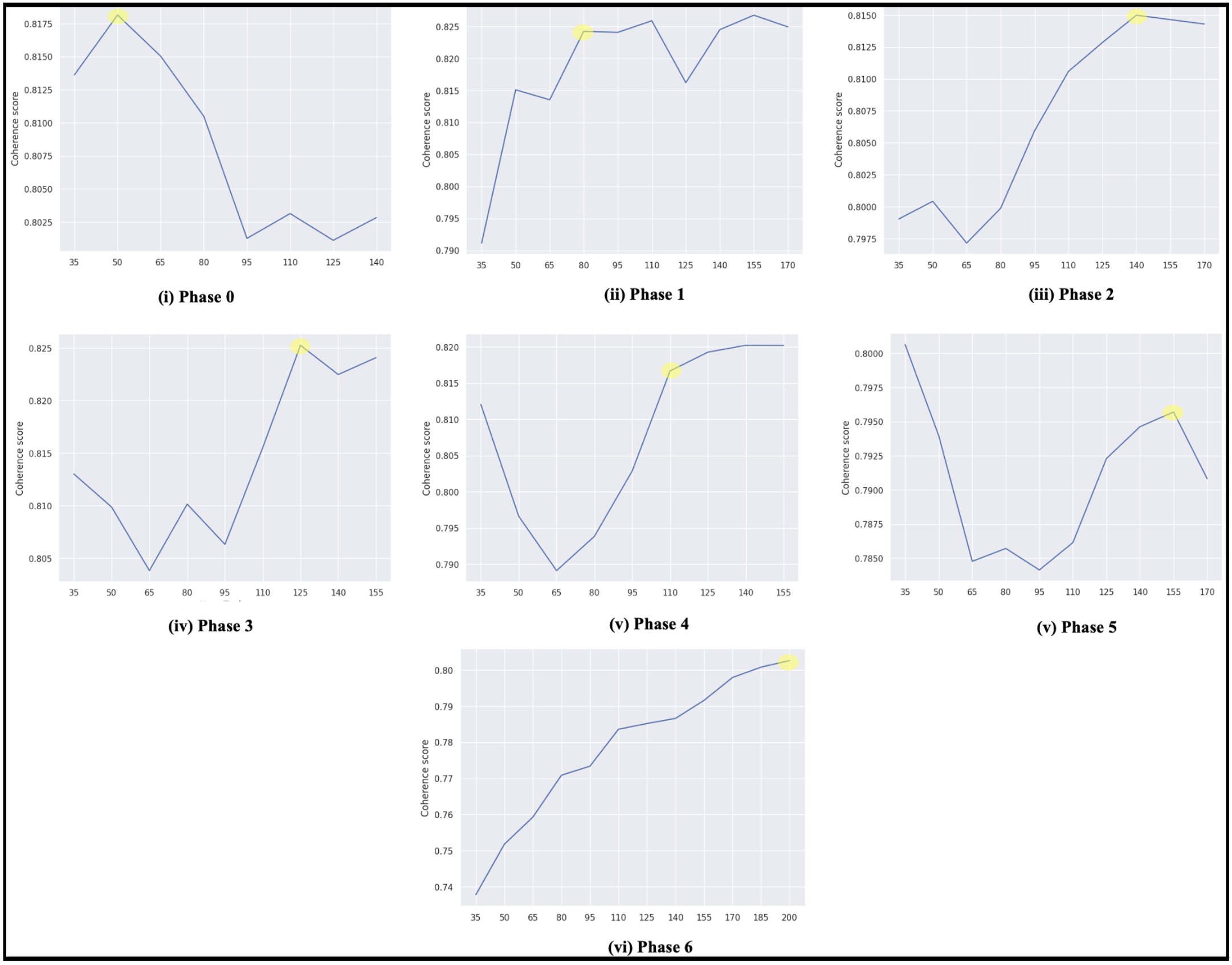}
    \caption{Coherence scores with optimal number of topics for BERTopic models across different phases during the conflict. The average number of topics ranges between 50 and 150 for all phases except Phase 6. The highlighted points represent the selected number of topics for their respective phase.}
    \label{fig:coherence_score}
    \label{Appendix: Figure B2}
\end{figure}

\begin{figure}[H]
    \centering
    \includegraphics[width=\textwidth]{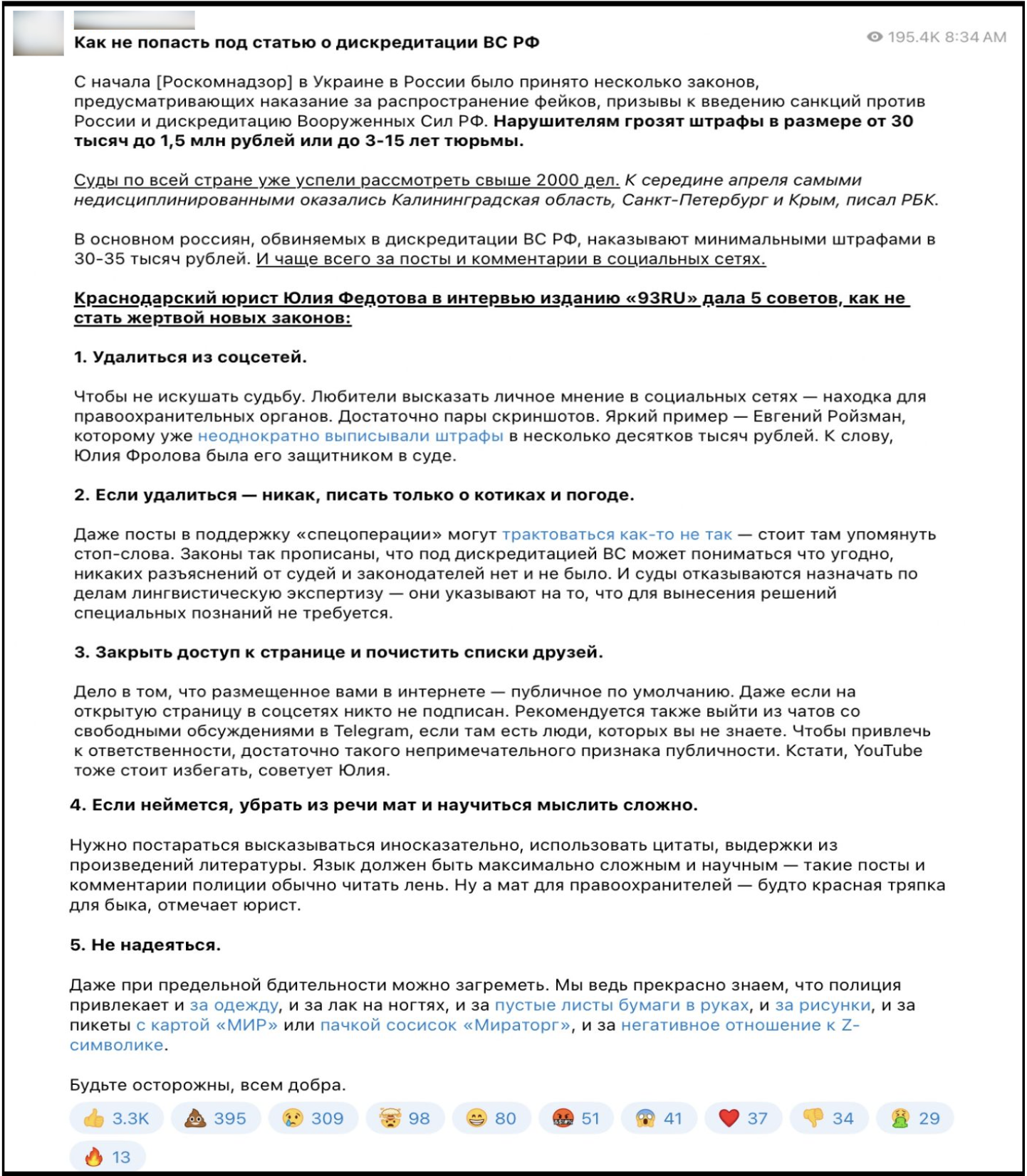}
    \caption{Post describing “How to avoid getting caught under an article about discrediting the RF Armed Forces”. Post garnered 195.4k views.}
    \label{Appendix: Figure B3}
\end{figure}

\begin{figure}[H]
    \centering
    \includegraphics[width=\textwidth]{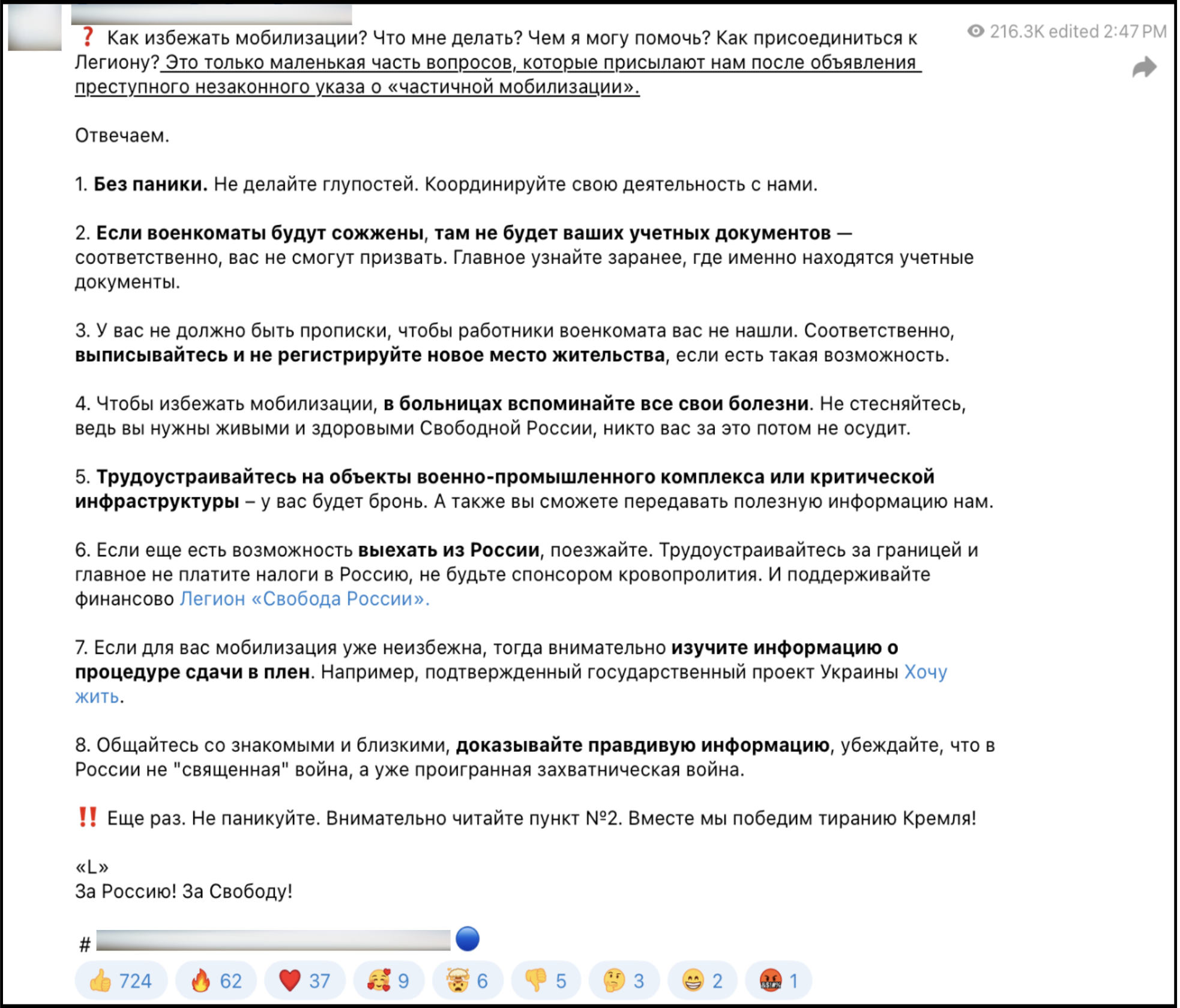}
    \caption{Post describing “How to avoid mobilization”. Post garnered 216.3k views.}
    \label{Appendix: Figure B4}
\end{figure}

\begin{figure}[H]
    \centering
    \includegraphics[width=\textwidth]{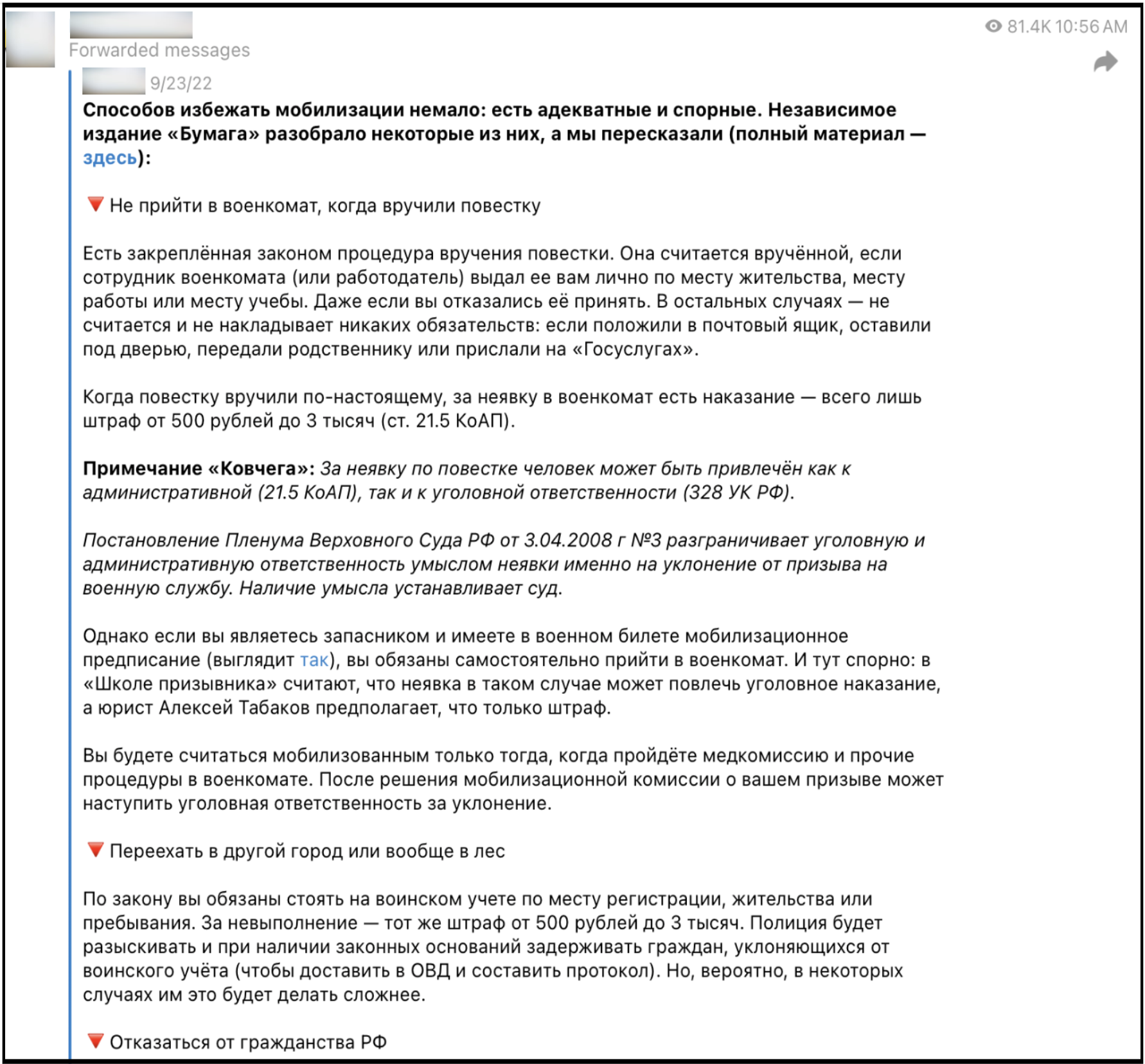}
    \caption{Another post on “How to avoid mobilization”. Post garnered 81.4k views.}
    \label{Appendix: Figure B5}
\end{figure}

\begin{figure}[H]
    \centering
    \includegraphics[width=\textwidth]{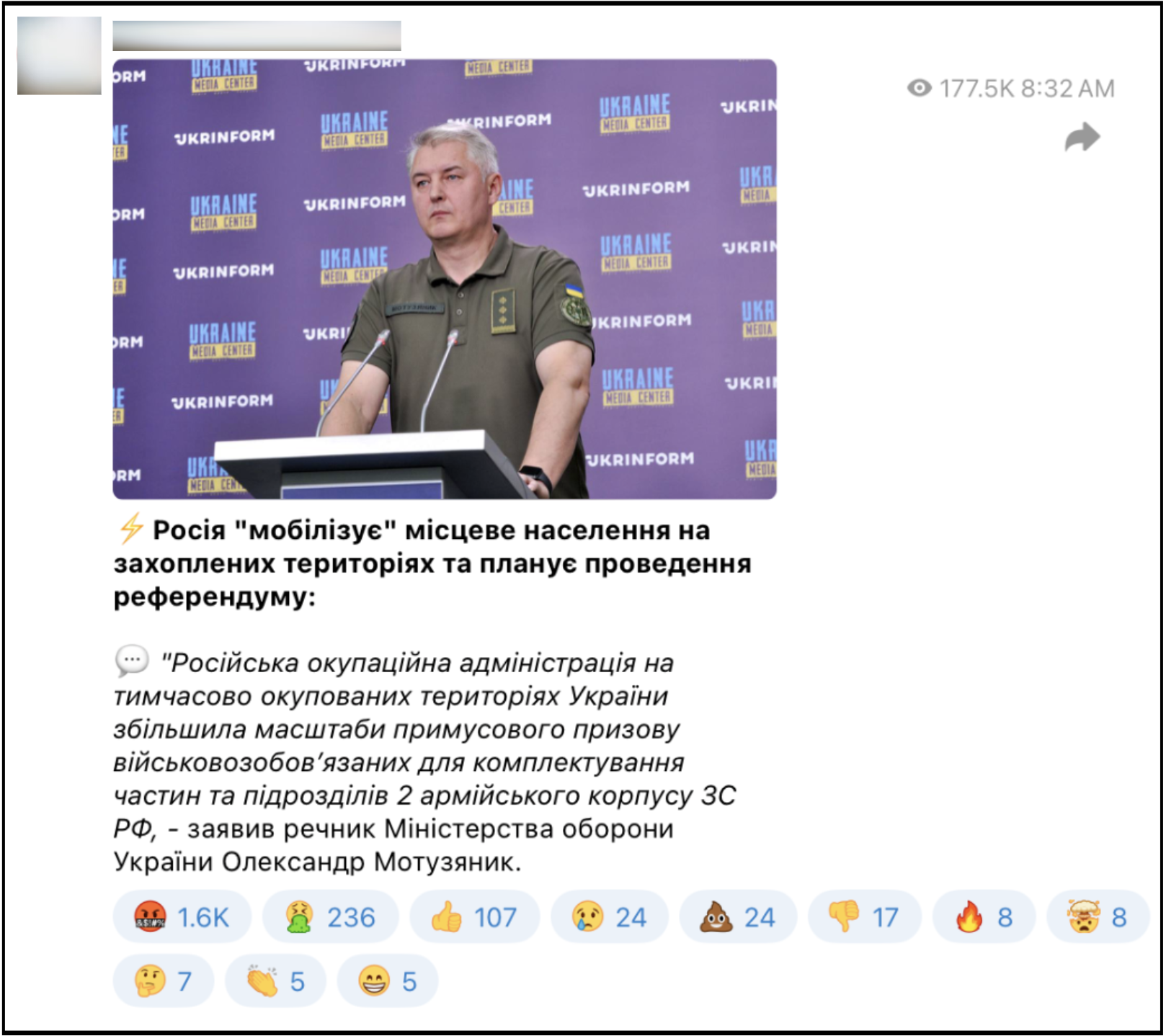}
    \caption{Post on “Russia mobilizes the local population in the captured territory and plans to hold a referendum”. Post garnered mostly negative reactions and was viewed 177.5k times.}
    \label{Appendix: Figure B6}
\end{figure}

\begin{figure}[H]
    \centering
    \includegraphics[width=\textwidth]{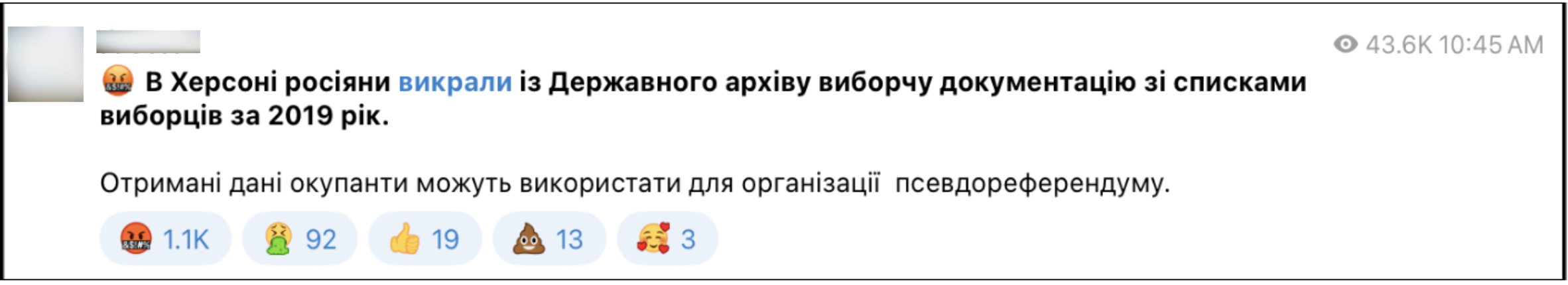}
    \caption{Post on stealing election documents to organize a pseudo referendum. Post garnered 43.6k views.}
    \label{Appendix: Figure B7}
\end{figure}

\begin{figure}[H]
    \centering
    \includegraphics[width=\textwidth]{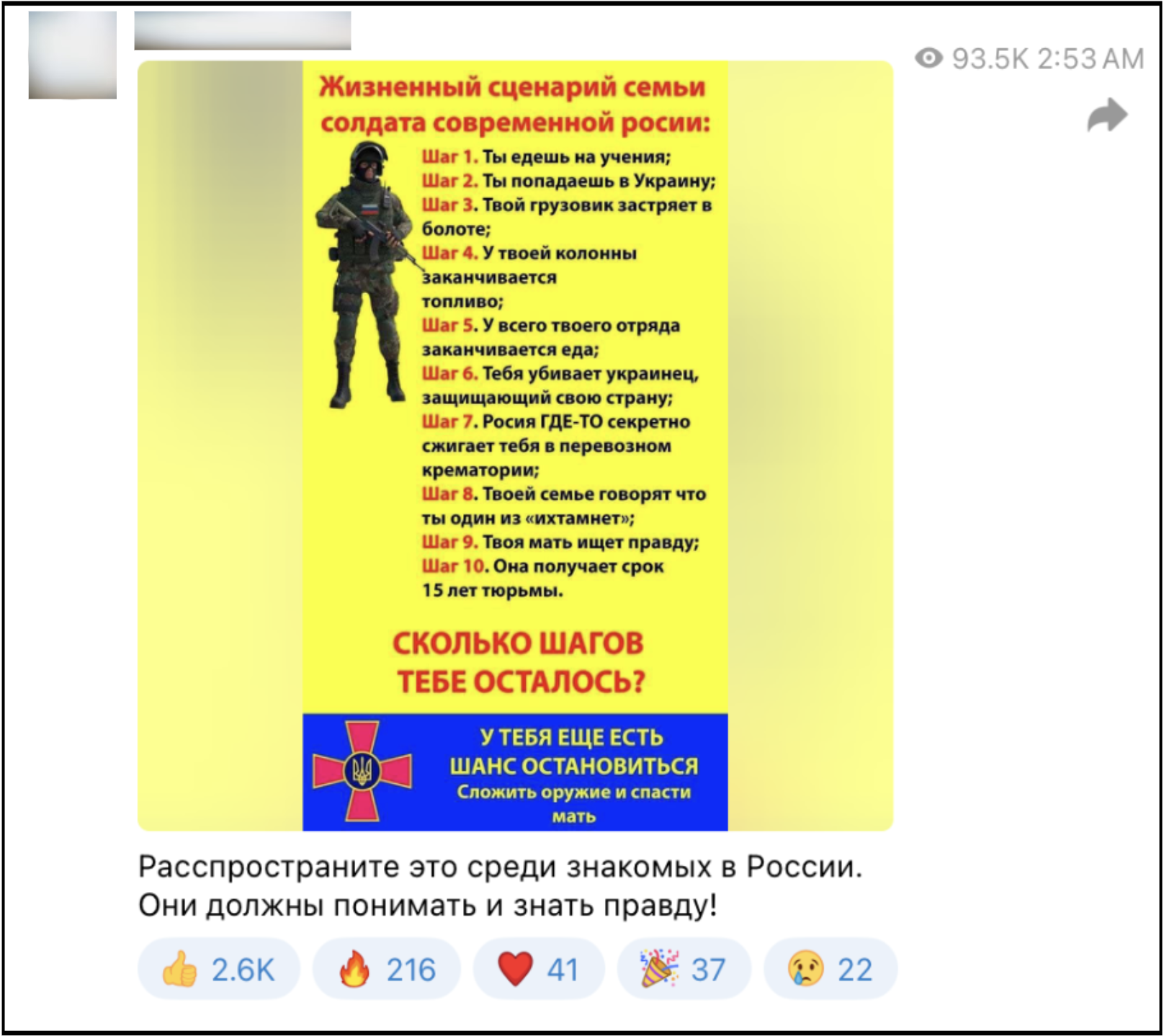}
    \caption{Post on insisting soldiers to lay down their weapons. Post garnered 93.5k views.}
    \label{Appendix: Figure B8}
\end{figure}

\section{Offline Events Timeline}\label{secA3} \label{Appendix C}

\subsection*{Phase 1: February 24 to March 30 — Full Invasion}

\noindent\textbf{February 2022}

\begin{itemize}
    \item \textbf{February 18}: Russia may launch an attack on Ukraine. The attack would likely begin with an air and missile campaign targeting much of Ukraine to decapitate the government and degrade the Ukrainian military as well as the ability of Ukrainian citizens to prepare to resist a subsequent Russian invasion.
    
    \item \textbf{February 21}: Russia recognized the Donetsk and Luhansk People’s Republics (DNR and LNR) and is deploying troops to Donetsk and Luhansk on the night of February 21, 2022.
    
    \item \textbf{February 22}: The US and its European allies defined Putin’s recognition of the DNR and LNR as an invasion of Ukraine and imposed a first round of sanctions. The Russian stock market and Ruble plummeted as the Kremlin sought to reassure Russia’s population that Russia could weather Western sanctions.
    
    \item \textbf{February 24}: Russian President Vladimir Putin began a large-scale invasion of Ukraine.
    
    \item \textbf{February 25}: Russian forces entered major Ukrainian cities—including Kyiv and Kherson—for the first time. NATO activated its 40,000-troop response force. The West removed select Russian banks from the SWIFT global financial network.
\end{itemize}

\noindent\textbf{March 2022}

\begin{itemize}
    \item \textbf{March 3}: Russian troops have surrounded Mariupol and are attacking it brutally to compel its capitulation or destroy it. Georgia and Moldova officially applied to join the European Union.
    
    \item \textbf{March 4}: NATO rejected Ukraine’s request to establish a no-fly zone over Ukraine.
    
    \item \textbf{March 11}: The Kremlin announced plans to deploy foreign fighters, including up to 16,000 Syrian fighters, to Ukraine.
    
    \item \textbf{March 14}: Russia and China deny that Russia seeks military aid from China.
    
    \item \textbf{March 16}: Ukrainian forces shot down 10 Russian aircraft—including five jets, three helicopters, and two UAVs. Russian President Vladimir Putin asked China for military and economic support for the war in Ukraine. China has neither confirmed nor denied whether they will provide aid to Russia.
\end{itemize}

\subsection*{Phase 2: April 1 to May 11 — Western Boost}

\noindent\textbf{April 2022}

\begin{itemize}
    \item \textbf{April 1}: Bucha atrocities uncovered. When Russian troops withdrew from Bucha in early April, they left behind a trail of destruction — and evidence of summary executions, brutality, and indiscriminate shelling.
    
    \item \textbf{April 3}: Ukraine has won the Battle of Kyiv, and Russian forces are completing their withdrawals from both the east and the west banks of the Dnipro in disorder.
    
    \item \textbf{April 14}: The flagship of Russia’s Black Sea Fleet sank following a likely Ukrainian cruise missile strike.
\end{itemize}

\noindent\textbf{May 2022}

\begin{itemize}
    \item \textbf{May 5}: Sweden and Finland are considering NATO membership.
    
    \item \textbf{May 9}: Russian President Vladimir Putin used his Victory Day speech to praise ongoing Russian efforts in Ukraine and reinforce existing Kremlin framing rather than announcing a change.
    
    \item \textbf{May 13}: Ukraine has likely won the Battle of Kharkiv. Russian forces continued to withdraw from the northern settlements around Kharkiv City.
    
    \item \textbf{May 17}: The Ukrainian military command ordered the defenders of Azovstal steel plant, Mariupol, to surrender.
\end{itemize}

\subsection*{Phase 3: June 1 to July 31 — Attrition/Stalemate}

\noindent\textbf{June 2022}

\begin{itemize}
    \item \textbf{June 3}: Russian occupation authorities began issuing Russian passports in Kherson City.
    
    \item \textbf{June 20}: Ukrainian sources confirmed that Russian forces control Severodonetsk city with the exception of the Azot industrial zone.
    
    \item \textbf{June 26}: Russian forces conducted a missile strike against Kyiv for the first time since April 29, likely to coincide with the ongoing G7 leadership summit.
    
    \item \textbf{June 27}: Russia defaulted on its foreign-currency sovereign debt for the first time in a century.
\end{itemize}

\noindent\textbf{July 2022}

\begin{itemize}
    \item \textbf{July 3}: Russian forces capture the city of Lysychansk in Eastern Ukraine.
\end{itemize}

\subsection*{Phase 4: August 1 to October 7 — Ukrainian Counteroffensive}

\noindent\textbf{August 2022}

\begin{itemize}
    \item \textbf{August 9}: Ukraine attacked the Saki Air Base in Russian-occupied Crimea, over 225 km behind Russian lines, which destroyed at least eight Russian aircraft and multiple buildings.
    
    \item \textbf{August 12}: Ukrainian forces destroyed the last functioning bridge Russian forces used to transport military equipment near the Kakhovka Hydroelectric Power Plant.
    
    \item \textbf{August 23}: Russian government sources confirmed that Russian authorities are bringing Ukrainian children to Russia and having Russian families adopt them.
\end{itemize}

\noindent\textbf{September 2022}

\begin{itemize}
    \item \textbf{September 1}: European gas prices spike by as much as 30\% after Russia says one of its main gas supply pipelines to Europe will remain closed indefinitely.
    
    \item \textbf{September 5}: Power unit No. 6 of the ZNPP became disconnected from the Ukrainian power grid.
    
    \item \textbf{September 6-7}: Ukrainian forces gained 400 sq km of territory northwest of Izyum as part of a highly effective counteroffensive in southeastern Kharkiv Oblast.
    
    \item \textbf{September 13}: The Kremlin has recognized its defeat in Kharkiv Oblast, the first defeat Russia has acknowledged in this war.
    
    \item \textbf{September 14}: Wagner Group financier Yevgeny Prigozhin is being established as the face of the Russian “special military operation” in Ukraine.
    
    \item \textbf{September 21}: Putin delivered a speech outlining his plan to mobilize an additional 300,000 troops in an effort to reclaim lost territory.
    
    \item \textbf{September 30}: Russian President Vladimir Putin announced the Russian annexation of four Ukrainian territories without clearly defining the borders of claimed territories.
\end{itemize}

\subsection*{Phase 5: October 8 to December — Russia’s Escalation}

\noindent\textbf{October 2022}

\begin{itemize}
    \item \textbf{October 8}: A large-scale explosion damaged the Kerch Strait Bridge that links occupied Crimea with Russia.
    
    \item \textbf{October 10-11}: Russian forces conducted massive, coordinated missile strikes on over 20 Ukrainian cities.
    
    \item \textbf{October 31}: Russian forces launched another massive wave of strikes against critical Ukrainian infrastructure, further damaging the power grid and leaving much of Kyiv without water.
\end{itemize}

\noindent\textbf{November 2022}

\begin{itemize}
    \item \textbf{November 5}: Iranian Foreign Minister Hossein Amir-Abdollahian confirmed that Iran began providing Russia drones.
    
    \item \textbf{November 12}: Kherson city was liberated.
    
    \item \textbf{November 13}: President Putin proposed an amendment to a draft law that would allow Russian officials to revoke Russian citizenship for disseminating “false” information about the Russian military, participating in extremist or undesirable organizations, or calling for violations of Russian “territorial integrity.”
    
    \item \textbf{November 15}: Russian forces conducted the largest set of missile strikes against Ukrainian critical infrastructure since the start of the war. Polish officials announced that a likely “Russian-made missile” landed in Poland within six kilometers of the international border with Ukraine.
\end{itemize}

\noindent\textbf{December 2022}

\begin{itemize}
    \item \textbf{December 21}: Ukrainian President Volodymyr Zelensky traveled to the USA.
    
    \item \textbf{December 29}: Russian forces conducted another massive series of missile strikes against Ukrainian critical infrastructure.
\end{itemize}

\subsection*{Phase 6: January 1 to March — Advanced Weapons Support for Ukraine}

\noindent\textbf{January 2023}

\begin{itemize}
    \item \textbf{January 5}: Russian President Vladimir Putin’s announcement that Russian forces will conduct a 36-hour ceasefire between January 6 and January 7 in observance of Russian Orthodox Christmas.
    
    \item \textbf{January 24}: A coalition of NATO member states reportedly will send Ukraine modern main battle tanks.
\end{itemize}

\noindent\textbf{February 2023}

\begin{itemize}
    \item \textbf{February 24}: US President Joe Biden visited Kyiv ahead of the first anniversary of Russia’s invasion of Ukraine. Western governments made a variety of statements on the provision of military aid to Ukraine.
    
    \item \textbf{February 25}: US President Joe Biden rejected China’s 12-point peace plan.
\end{itemize}

\noindent\textbf{March 2023}

\begin{itemize}
    \item \textbf{March 1}: Belarusian President Lukashenko and Chinese President Xi Jinping signed a package that may facilitate Russian sanctions evasion by channelling Chinese aid to Russia through Belarus.
    
    \item \textbf{March 8}: German and Polish officials announced that Germany and Poland will deliver 28 Leopard 2 tanks to Ukraine in March 2023, which will bolster Ukraine’s capabilities to conduct a counteroffensive amidst high Russian tank losses.
    
    \item \textbf{March 12}: Iranian State Media announced on March 11 that Iran has finalized a deal to buy Sukhoi-35 fighter jets from Russia.
    
    \item \textbf{March 20}: Chinese President Xi Jinping met with Russian President Vladimir Putin in Moscow.
    
    \item \textbf{March 21}: US Department of Defense (DoD) announced \$350 million of security assistance to Ukraine.
    
    \item \textbf{March 29}: Iranian Foreign Affairs Minister Hossein Amir Abdollahian met with Russian Foreign Minister Sergei Lavrov in Moscow.
\end{itemize}

\end{appendices}

\end{document}